\documentstyle[aps,prb,floats,epsf,psfig,12pt]{revtex}

\tightenlines
\input advcp.def
\input{psfig}
 
\gdef\journal#1, #2, #3, 1#4#5#6{		
    {\rm #1~}{\bf #2}, #3 (1#4#5#6)}		

\def\bio{\journal Biometrika, }

\def\commpureapplmath{\journal Comm. Pure Appl. Math., }

\def\cpl{\journal Chem. Phys. Lett., }

\def\etal{{\it et al.}}

\def\jcompp{\journal J. Comp. Phys., }
\def\jcp{\journal J. Chem. Phys., }

\def\jsp{\journal J. Stat. Phys., }

\def\physrep{\journal Phys. Reports, }
\def\physrev{\journal Phys. Rev., }

\def\pra{\journal Phys. Rev. A, }
\def\prb{\journal Phys. Rev. B, }

\def\prl{\journal Phys. Rev. Lett., }

\def\statphys{\journal J. Stat. Phys., }

\begin{document}

\title{Monte Carlo Eigenvalue Methods in Quantum
Mechanics and Statistical Mechanics
\footnote{
Advances in Chemical Physics, Vol. 105, Monte Carlo Methods in Chemistry,
edited by David M. Ferguson, J. Ilja Siepmann, and Donald G. Truhlar,
series editors I. Prigogine and Stuart A. Rice, Chapter 4
(In press, Wiley, NY 1998)}
}

\author{M. P. Nightingale}
\address{Department of Physics, University of Rhode Island,\\
Kingston, RI 02881.}

\author{C. J. Umrigar}
\address{Cornell Theory Center and Laboratory of Atomic
and Solid State Physics,\\
Cornell University, Ithaca, NY 14853.}

\date{\today}

\maketitle
\begin{abstract}

In this review we discuss, from a unified point of view, a variety of
Monte Carlo methods used to solve eigenvalue problems in statistical
mechanics and quantum mechanics.  Although the applications of these
methods differ widely, the underlying mathematics is quite similar in
that they are stochastic implementations of the power method.  In all
cases, optimized trial states can be used
to reduce the errors of Monte Carlo estimates.

\end{abstract}

\tableofcontents

\section{Introduction}

Many important problems in computational physics and chemistry can be
reduced to the computation of dominant eigenvalues of matrices of high
or infinite order.  We shall focus on just a few of the numerous
examples of such matrices, namely, quantum mechanical Hamiltonians,
Markov matrices and transfer matrices.  Quantum Hamiltonians, unlike
the other two, probably can do without introduction.  Markov matrices
are used both in equilibrium and nonequilibrium statistical mechanics
to describe dynamical phenomena.  Transfer matrices were introduced by
Kramers and Wannier in 1941 to study the two-dimensional Ising model,
\cite{KraWa} and ever since, important work on lattice models in
classical statistical mechanics has been done with transfer matrices,
producing both exact and numerical results.\cite{TMreview}

The basic \MC\ methods reviewed in this chapter have been used in many
different contexts and under many different names for many decades,
but we emphasize the solution of eigenvalue problems by means of \MC\
methods and present the methods from a unified point of view.  A vital
ingredient in the methods discussed here is the use of optimized trial
functions.  Section~\ref{sec.optimiz} deals with this topic briefly,
but in general we suppose that optimized trial functions are given.
We refer the reader to Ref.~\onlinecite{NU96} for more details on
their construction.

The analogy of the time-evolution operator in quantum mechanics on the
one hand, and the transfer matrix and the Markov matrix in statistical
mechanics on the other, allows the two fields to share numerous
techniques.  Specifically, a transfer matrix $G$ of a statistical
mechanical lattice system in $d$ dimensions often can be interpreted
as the evolution operator in discrete, imaginary time $t$ of a quantum
mechanical analog in $d-1$ dimensions.  That is, $
G\approx\exp(-t{\cal H})$, where $\cal H$ is the Hamiltonian of a
system in $d-1$ dimensions, the quantum mechanical analog of the
statistical mechanical system.  From this point of view, the
computation of the partition function and of the ground-state energy
are essentially the same problems: finding the largest eigenvalue of
$G$ and of $\exp(-t{\cal H})$, respectively.  As far as the Markov
matrix is concerned, this simply is the time-evolution operator of a
system evolving according to stochastic dynamics.  The largest
eigenvalue of such matrices equals unity, as follows from conservation of probability,
and for systems in thermal equilibrium, the corresponding eigenstate is
also known, namely the Boltzmann distribution.  Clearly, the dominant
eigenstate in this case poses no problem.  For nonequilibrium
systems, the stationary state is unknown and one might use the methods
described in this chapter in dealing with them.  Another problem is the
computation of the relaxation time of a system with stochastic
dynamics.  This problem is equivalent to the computation of the second
largest eigenvalue of the Markov matrix, and again the current methods
apply.

The emphasis of this chapter is on methods rather than applications, but
the reader should have a general idea of the kind of problems for
which these methods can be employed.  Therefore, we start off by
giving some specific examples of the physical systems one can deal
with.

\subsection{Quantum Systems}
In the case of a quantum mechanical system, the problem in general is
to compute expectation values, in particular the energy, of
bosonic or fermionic ground or excited eigenstates.
For systems with $n$
electrons, the spatial coordinates are denoted by a $3n$-dimensional
vector $\R$. In terms of the vectors ${\bf r}_i$ specifying the
coordinates of electron number $i$ this reads $\R=({\bf r}_1,\dots,{\bf
r}_n)$.  The dimensionless Hamiltonian is of the form
\beq
\bra{\R}|\Ham \ket{\Rp} =
[-{1 \over 2 \mu}\Grad^2 + \V(\R)] \delta(\R-\Rp).
\label{eq.hamiltonian}
\eeq
For atoms or molecules atomic units are used $\mu=1$, and $\V$ is the
usual Coulomb potential acting between the electrons and between the
electrons and nuclei {\it i.e.,}
\beq
\V(\R) = \sum_{i<j} {1 \over r_{ij}} -
\sum_{\alpha,i} {Z_\alpha \over r_{\alpha i}}
\label{eq.Coulomb}
\eeq
where for arbitrary subscripts $a$ and $b$ we define
$r_{ab} = |{\bf r}_a - {\bf r}_b|$\,;
indices $i$ and $j$ label the electrons, and we assume that the nuclei are
of infinite mass and that nucleus $\alpha$ has charge $Z_\alpha$ and
is located at position ${\bf r}_\alpha$.

In the case of quantum mechanical van der Waals
clusters,\cite{MN94,MMN96} $\mu$ is the reduced mass --- $\mu=
2^{1\over 3} m \epsilon \sigma /\hbar^2$ in terms of the mass $m$,
Planck's constant $\hbar$ and the conventional Lennard-Jones parameters
$\epsilon$ and $\sigma$ --- and the potential is given by
\beq
\V(\R)=\sum_{i<j} {1 \over r^{12}_{ij}} - {2 \over r^{6}_{ij}}
\label{eq.LennardJones}
\eeq
The quantum nature of the system increases with $1/\mu^2$, which is
proportional to the conventional de Boer parameter.

The ground-state wavefunction of a bosonic system is positive
everywhere, which is very convenient in a Monte Carlo context and
allows one to obtain results with an accuracy that is limited only by
practical considerations.  For fermionic systems, the ground-state wave
function has nodes, and this places more fundamental limits on the
accuracy one can obtain with reasonable effort.  In the methods
discussed in this chapter, this bound on the accuracy takes the form of
the so-called {\em fixed-node approximation.}  Here one assumes that
the nodal surface is given, and computes the ground-state wave
function subject to this constraint.

The time-evolution operator $\exp(-\tau \Ham)$ in the position
representation is the Green function
\beq
\Grprt = \bra \Rvecp | e^{-\tau \Ham} \ket{\Rvec}.
\eeq
For both bosonic systems and fermionic systems in the fixed-node
approximation, $G$ has only nonnegative elements.
This is essential for the \MC\
methods discussed here.  A problem specific to quantum mechanical
systems is that $G$ is only known asymptotically for short times, so
that the finite-time Green function has to
be constructed by the application of the generalized Trotter
formula\cite{Suzuki76,Suzuki77}, \ie, $G(\tau)=\lim_{m \to
\infty} G(\tau/m)^m$, where the position variables of $G$ have been
suppressed.

\subsection{Transfer Matrices}
\label{sec.transfer}
Our next example is the transfer matrix of statistical mechanics.  The
largest eigenvalue yields the free energy, from which all
thermodynamic properties follow.  As a typical transfer matrix, one can
think of the one-site, Kramers-Wannier transfer matrix for a
two-dimensional model of Ising spins, $s_i=\pm 1$.  Such a matrix
takes a particularly simple form for a square lattice wrapped on a
cylinder with helical boundary conditions with pitch one.  This
produces a mismatch of one lattice site for a path on the lattice
around the cylinder.  This geometry has the advantage that a
two-dimensional lattice can be built one site at a time and that the
process of adding each single site is identical each time.  Suppose we
choose a lattice of $M$ sites, wrapped on a cylinder with a
circumference of $L$ lattice spaces.  Imagine that we are adding sites
so that the lattice grows toward the left. We can then define a
conditional partition function $Z_M(\S)$, which is a sum over those
states (also referred to as configurations) for which the left-most edge
of the lattice is in a given state $\S$.
The physical interpretation of $Z_M(\S)$ is the relative
probability of finding the left-most edge in a state $\S$ with which
the rest of the lattice to its right is in thermal equilibrium.

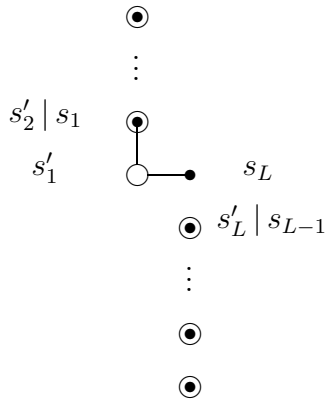
\begin{figure}

\begin{center}
\setlength{\unitlength}{2pt}
\begin{picture}(20,75)(-5,-20)
\put( 0,30){\circle{4}}
\put(-20,30){$s'_1$}
\put( 0,40){\circle{4}}
\put( 0,40){\circle*{2}}
\put(-28,40){$\ \;s'_{2}\,|\,s_{1}$}
\put(-1,48){\vdots}
\put( 0,60){\circle{4}}
\put( 0,60){\circle*{2}}
\put(10, 0){\circle{4}}
\put(10, 0){\circle*{2}}
\put(10,-10){\circle{4}}
\put(10,-10){\circle*{2}}
\put( 9, 8){\vdots}
\put(10,20){\circle{4}}
\put(10,20){\circle*{2}}
\put(15,20){$s'_L\,|\,s_{L-1}$}
\put(10,30){\circle*{2}}
\put(15,30){$\,\ \ s_L$}
\put( 0,32){\line(0,7)7}
\put( 2,30){\line(7,0)7}
\end{picture}
\end{center}

\caption[c1]{ \footnotesize

Graphical representation of the transfer matrix. The primed variables
are associated with the circles and combine into the left index of
the matrix; the dots go with the right index and the unprimed
variables.  Coincidence of a circle and a dot produces a
$\delta$-function.  An edge indicates a contribution to the Boltzmann
weight.  Repeated application of this matrix constructs a lattice with
nearest neighbor bonds and helical boundary conditions.}

\label{fig.transfer_matrix}
\end{figure}

If one has helical boundary conditions and spins that interact only
with their nearest neighbors, one can repeatedly add just a single
site and the bonds connecting it to its neighbors above and to the
right.  Analogously, the transfer matrix $G$ can be used to compute
recursively the conditional partition function of a lattice with one
additional site
\begin{equation}
Z_{M+1}(\Sp)=\sum_{\S} G(\Sp|\S) Z_{M}(\S),
\label{eq.Z}
\end{equation}
with
\begin{equation}
G(\Sp|\S) =
\exp[K(s'_1 s_1 +  s'_1 s_L)]
\prod_{i=2}^{L} \delta_{s'_i, s_{i-1}},
\label{eq.tm}
\end{equation}
with $\S =(s_1,s_2,\dots,s_L)$ and $\Sp =(s'_1,s'_2,\dots,s'_L)$, and
the $s_i,s'_i=\pm 1$ are Ising spins.  With this definition of the
transfer matrix, the matrix multiplication in Eq.~(\ref{eq.Z})
accomplishes the following: (1) a new site, labeled 1, is appended to
the lattice at the left edge; (2) the Boltzmann weight is updated to
that of the lattice with increased size; (3) the old site $L$ is
thermalized; and finally (4) old sites $1,\dots,L-1$ are pushed down
on the stack and are renamed to $2,\dots,L$.
Sites have to remain in the stack until all interacting
sites have been added, which determines the minimal depth of the
stack.  It is clear from Figure \ref{fig.transfer_matrix} that the
transfer matrix is nonsymmetric and indeed symmetry is not required
for the methods discussed in this chapter. It is of some interest that
transfer matrices usually have the property that a right eigenvector
can be transformed into a left eigenvector by a simple permutation and
reweighting transformation.  The details are not important here and
let it suffice to mention that this follows from an obvious symmetry
of the diagram shown in Figure~\ref{fig.transfer_matrix}: (1) rotate
over $\pi$ about an axis perpendicular to the paper, which permutes
the states; and (2) move the vertical bond back to its original
position, which amounts to reweighting by the Boltzmann weight of a
single bond.

Equation~(\ref{eq.Z}) implies that for large $M$ and generic boundary
conditions at the right-hand edge of the lattice, the partition
function approaches a dominant right eigenvector $\psi_0$ of the
transfer matrix $G$
\beq
Z_M(\S) \propto \lambda_0^M \psi_0(\S),
\eeq
where $\lambda_0$ is the dominant eigenvalue.  Consequently, for $M
\to\infty$ the free energy per site is given by
\beq
f=-kT \ln  \lambda_0.
\label{eq.free}
\eeq
The problem relevant to this chapter is the computation of the
eigenvalue $\lambda_0$ by \MC.\cite{NB96.prb}


\subsection{Markov Matrices}
Discrete-time Markov processes are a third type of problem we shall
discuss.  One of the challenges in this case is to compute the
correlation time of such a process in the vicinity of a critical
point, where the correlation time goes to infinity, a phenomenon
called ``critical slowing down".  Computationally, the problem amounts to
the evaluation of the second largest eigenvalue of the Markov matrix,
or more precisely its difference from unity. The latter goes to zero as the
correlation time approaches infinity.

The Markov matrix defines the stochastic evolution of the system in
discrete time.  That is, suppose that at time $t$ the probability of
finding the system in state $\S$ is given by $\rho_t (\S)$.  If the
probability of making a transition from state $\S$ to state $\Sp$ is
$\Phat(\Sp|\S)$ (sorry about the hat, we shall take it off soon!),
then
\beq
\rho_{t+1}(\Sp) = \sum_{\S} \Phat(\Sp|\S) \rho_t(\S)
\eeq
In the case of interest here, the Markov matrix $\Phat$ is constructed
so that its stationary state is the Boltzmann distribution
$\psiB^2=\exp( -\beta \Ham)$.  Sufficient conditions are that (a) each
state can be reached from every state in a finite number of
transitions and that (b) $\Phat$ satisfies detailed balance
\beq
\Phat(\Sp| \S) \psiB(\S)^2 = \Phat(\S|\Sp) \psiB(\Sp)^2.
\eeq
It immediately follows from detailed balance that the matrix $\Phat$
defined by
\beq
P(\Sp|\S) = {1\over \psiB(\Sp)}\Phat(\Sp|\S)\psiB(\S).
\eeq
is symmetric and equivalent by similarity transformation to the
original Markov matrix. Because of this symmetry,
expressions tend to take a simpler form when
$P$ is used, as we shall do, but one should keep in mind that
$P$ itself is not a Markov matrix, since the sum on its left index
does not yield unity identically.

Again, to provide a specific example, we mention that the methods
discussed below have been applied\cite{NB96.prl,NB97.prl} to an Ising
model on a square lattice with the heat bath or Yang~\cite{Yang}
transition probabilities and random site selection. In that case, time
evolution takes place by single spin-flip transitions which occur
between a given state $\S$ to one of the states $\Sp$ that differ only
at one randomly selected site.  For any such pair of states, the
transition probability is given by
\begin{equation}
\Phat(\Sp | \S ) = \left\{ \begin{array}{ll}
{1\over 2 N} {e^{-{1 \over 2}\beta \Delta \Ham}
\over\cosh {1\over 2}\beta\Delta
\Ham} &{\rm for} \;\; \Sp \ne \S \\
1-\sum_{\Spp \ne \S} \Phat(\Spp|\S) &{\rm for} \;\; \Sp = \S,
\end{array}
\right.
\end{equation}
for a system of $N$ sites with Hamiltonian
\beq
-\beta \Ham = K \sum_{(i,j)} s_i s_j + K' \sum_{(i,j)'} s_i s_j \; ,
\eeq
where $(i,j)$ denotes nearest-neighbor pairs,
and $(i,j)'$ denotes next-nearest-neighbor pairs and
$\Delta \Ham \equiv \Ham(\Sp)-\Ham(\S)$.

We note that whereas the transfer matrix is used to deal with systems
that are infinite in one direction, the systems used in the dynamical
computations are of finite spatial dimensions only.

\section{The Power Method}
\label{section.powermethod}
Before discussing technical details of Monte Carlo methods to compute
eigenvalues and expectation values, we introduce the mathematical
ideas and the types of expressions for which statistical estimates are
sought. We formulate the problem in terms of an operator $G$ of which
one wants to compute the dominant eigenstate and eigenvalue,
$\ket{\psi_0}$ and $\lambda_0$.  Mathematically, but not necessarily
in a \MC\ setting, dominant may mean dominant relative to eigenstates
of a given symmetry only.

The methods to be discussed are variations of the power method, which
relies on the fact that for a generic initial state $\ket {u^{(0)}}$
of the appropriate symmetry, the states $\ket {u^{(t)}}$ defined by
\beq
\ket {u^{(t+1)}} = {1\over c_{t+1}}G \ket {u^{(t)}}
\label{eq.power.method}
\eeq
converge to the dominant eigenstate $\ket{\psi_0}$ of $G$, if the
constants $c_t$ are chosen so that $\ket {u^{(t)}}$ assumes a standard
form, in which case the constants $c_t$ converge to the dominant
eigenvalue.  This follows immediately by expanding the initial state
$\ket {u^{(0)}}$ in eigenstates of $G$. One possible standard form is
that, in some convenient representation, the component of $\ket
{u^{(t)}}$ largest in magnitude equals unity.

For quantum mechanical systems, $G$ usually is the imaginary-time
evolution operator, $\exp (-\tau\Ham)$.  As mentioned above, a
technical problem in that case is that an explicit expression is known
only asymptotically for short times $\tau$.  In practice, this
asymptotic expression is used for a small but finite $\tau$ and this
leads to systematic, time-step errors.  We shall deal with this
problem below at length, but ignore it for the time being.

The exponential operator $\exp (-\tau\Ham)$ is one of various
alternatives that can be employed to compute the ground-state
properties of the Hamiltonian. If the latter is bounded from above,
one may be able to use $\openone -\tau\Ham$, where $\tau$ should be small enough
that $\lambda_0\equiv 1-\tau E_0$ is the dominant eigenvalue of
$\openone -\tau\Ham$.  In this case, there is no time-step error and
the same holds for yet another method of inverting the spectrum of the
Hamiltonian, {\it viz.} the Green function Monte Carlo method.  There
one uses $(\Ham-E)^{-1}$, where $E$ is a constant chosen so that the
ground state becomes the dominant eigenstate of this operator.  In a
Monte Carlo context, matrix elements of the respective operators are
proportional to transition probabilities and therefore have to be
non-negative, which, if one uses either of the last two methods, may
impose further restrictions on the values of $\tau$ and $E$.

For the statistical mechanical applications, the operators $G$ are
indeed evolution operators by construction.
The transfer matrix evolves the physical system in a spatial rather
than time direction, but this spatial direction corresponds to
time from the point of view of a \MC\ time series.  With this in mind,
we shall refer to the operator $G$ as the evolution operator, or the
{\em Monte Carlo} evolution operator, if it is necessary to
distinguish it from the usual time-evolution operator
$\exp (-\tau\Ham)$.

Suppose that $X$ is an operator of which one wants to compute an
expectation value. Particularly simple to deal with are the cases in
which the operators $X$ and $G$ are the same or commute.  We introduce
the following notation. Suppose that $\ket{u_\alpha}$ and
$\ket{u_\beta}$ are two states, then $X^{(p',p)}_{\alpha \beta}$
denotes the matrix element
\beq
X^{(p',p)}_{\alpha \beta} =
{\bra{u_\alpha}|G^{p'}XG^{p}\ket{u_\beta} \over
\bra{u_\alpha}|G^{p'+p}\ket{u_\beta}}.
\label{eq.general.estimator}
\eeq
This definition is chosen to simplify the discussion, and generalization to
physically relevant expressions, such as Eq.~\ref{eq.general.estimator2},
is straightforward.

Various Monte Carlo methods are designed to estimate particular
instances of $X^{(p',p)}_{\alpha \beta}$, and often the ultimate goal is to
compute the expectation value in the dominant eigenstate
\beq
X_0={\bra{\psi_0}|X\ket{\psi_0} \over \bra{\psi_0} \ket{\psi_0}},
\label{eq.X_0}
\eeq
which reduces to an expression for the dominant eigenvalue of interest
if one chooses for $X$, in the applications discussed in the
introduction, the Hamiltonian, transfer or Markov matrix.

The simplest method is the variational Monte Carlo method, discussed
in the next section.  Here an approximate expectation value is
computed by employing an approximate eigenvector of $G$.  Typically,
this is an optimized trial state, say $\ket{\trial{u}}$, in which case
variational Monte Carlo yields $X^{(0,0)}_{\TT}$, which is simply
the expectation value of $X$ in the trial state.  Clearly, variational
Monte Carlo estimates of $X_0$ have both systematic and statistical
errors.

The variational error can be removed asymptotically by projecting out
the dominant eigenstate, {\it i.e.,} by reducing the spectral weight
of sub-dominant eigenstates by means of the power method.  The
simplest case is obtained if one applies the power method only to the
right on the state $\ket{u_\beta}$ but not to the left on
$\bra{u_\alpha}|$ in Eq.~(\ref{eq.general.estimator}).
Mathematically, this is the essence of diffusion and transfer matrix
Monte Carlo, and in this way one obtains the desired result $X_0$ if
the operator $X$ commutes with the \MC\ evolution operator $G$.  In
our notation, this means that $X_0$ is given by the statistical
estimate of $X^{(0,\infty)}_{\TT}$. In principle, this yields an
unbiased estimate of $X_0$, but in practice one has to choose $p$ finite but
large enough that the estimated systematic error is less than the
statistical error.  In some practical situations it can in fact be
difficult to ascertain that this indeed is the case.  If one is
interested in the limiting behavior for infinite $p$ or $p'$, the
state $\ket{\ua}$ or $\ket{\ub}$ need not be available in closed form.
This freedom translates into the flexibility in algorithms design
exploited in diffusion and transfer matrix \MC.

If $G$ and $X$ do not commute, the mixed estimator
$X^{(0,\infty)}_{\TT}$ is not the desired result, but the residual
systematic error can be reduced by combining the variational and mixed
estimates by means of the expression
\beq
X_0=2 X_{\TT}^{(0,\infty)}-X_{\TT}^{(0,0)}+
{\cal O}[(\ket{\psi_0}-\ket{\trial{u}})^2]
\eeq

To remove the variational bias systematically, if $G$ and $X$ do not
commute, the power method must be used to both the left and the
right in Eq.~(\ref{eq.general.estimator}).  Thus one obtains from
$X^{(\infty,\infty)}_{\TT}$ an exact estimate of $X_0$ subject only to
statistical errors.  Of course, one has to pay the price of the
implied double limit in terms of loss of statistical accuracy.  In the
context of the Monte Carlo algorithms discussed below, such as
diffusion and transfer matrix Monte Carlo, this double projection
technique to estimate $X^{(\infty,\infty)}_{\TT}$ is called {\it
forward walking} or {\it future walking}.

We end this section on the power method with a brief discussion of the
computational complexity of using the \MC\ method for eigenvalue problems.
In \MC\ computations one can distinguish operations of three levels of
computational complexity, depending on whether the operations have to
do with single particles or lattice sites, the whole system, or state
space summation or integration. The first typically involves a fixed
number of elementary arithmetic operations, whereas this number
clearly is at least proportional to the system size in the second case.
Exhaustive state-space summation grows exponentially in the total
system size, and for these problems \MC\ is often the only
viable option.

Next, the convergence of the power method itself comes into play.  The
number of iterations required to reach a certain given accuracy is
proportional to $\log|\lambda_0/\lambda_1|$, where the $\lambda_0$ and
$\lambda_1$ are the eigenvalues of largest and second largest
magnitude.  If one is dealing with a single-site transfer matrix of a
critical system, that spectral gap is proportional to $L^{-d}$ for a
system in $d$ dimensions with a cross section of linear dimension $L$.
In this case, a single matrix multiplication is of the complexity of a
one-particle problem.  In contrast, both for the Markov matrix defined
above, and the quantum mechanical evolution operator, the matrix
multiplication itself is of system-size complexity.  Moreover, both of
these operators have their own specific problems.  The quantum
evolution operator of $G(\tau)$ has a gap on the order of $\tau$,
which means that $\tau$ should be chosen large for rapid convergence,
but one does not obtain the correct results, because of the time-step
error, unless $\tau$ is small.  Finally, the spectrum of the Markov
matrix displays critical slowing down.  This means that the gap
of the single spin-flip matrix is on the order of $L^{-d-z}$, where
$z$ is typically a little bigger than two.\cite{NB96.prl} These
convergence properties are well understood in terms of the mathematics
of the power method.  Not well understood, however, are problems that
are specific to the \MC\ implementation of this method, which in some
form or another introduces multiplicative fluctuating weights that are
correlated with the quantities of
interest.\cite{Hetherington,CeperleyBernu88}

\section{Single-Thread Monte Carlo}
\label{sec.singlethread}
In the previous section we have presented the mathematical expressions
that can be evaluated with the Monte Carlo algorithms to be discussed
next.  The first algorithm is designed to compute an approximate
statistical estimate of the matrix element $X_0$ by means of the
variational estimate $X_{\TT}^{(0,0)}$.
\renewcommand{\thefootnote}{\fnsymbol{footnote}}
We write\footnote{We assume throughout that the states we are dealing with are
represented by real numbers and that $X$ is represented by a real,
symmetric matrix.  In many cases, generalization to complex
numbers is trivial, but for some physical problems, while formal
generalization may still be possible, the resulting \MC\
algorithms may be too noisy to be practical.}
$\; \bra{\S}\ket{\uT}=\bra{\uT}\ket{\S}\equiv \uT(\S)$ and for
non-vanishing $\uT(\S)$ define the {\it configurational eigenvalue}
$\XT(\S)$ by
\beq
\XT(\S)\uT(\S) \equiv \bra{\S}| X\ket{\uT} =
\sum_{\Sp} \bra{\S}| X\ket{\Sp} \bra{\Sp}\ket{\uT}
\label{eq.XT}
\eeq
This yields
\beq
X_{\TT}^{(0,0)}=
{\sum_{\S}\uT(\S)^2 \XT(\S) \over {\sum_{\S} \uT(\S)^2}},
\label{eq.XTT.Srep}
\eeq
which shows that $X_{\TT}$ can be evaluated as a time average over a
Monte Carlo time series of states $\S_1,\S_2,\dots$ sampled from the
probability distribution $\uT(\S)^2$, {\it i.e.}, a process in which
$\mbox{Prob}(\S)$, the probability of finding a state $\S$ at any
time is given by
\beq
\mbox{Prob}(\S) \propto \uT(\S)^2.
\eeq
For such a process, the ensemble average in Eq.~(\ref{eq.XTT.Srep})
can be written in the form of a time average
\beq
X_{\TT}^{(0,0)}=\lim_{L\to\infty}{1\over L}
\sum_{t=1}^L\XT(\S_t).
\label{eq.X.MCav}
\eeq

For this to be of practical use, it has to be assumed that the
configurational eigenvalue $\XT(\S)$ can be computed efficiently,
which is the case if the sum over states $\S'$ in
$\bra{\S}|X\ket{\uT}=
\sum_{\S'} \bra {\S} | X \ket {\S'} \bra {\S'} \ket {\uT}$ can be
performed explicitly.  For discrete states this means that $X$ should
be represented by a sparse matrix; if the states $\S$ form a
continuum, $\XT(\S)$ can be computed directly if $X$ is diagonal or
{\it near-diagonal}, {\it i.e.,} involves no or only low-order derivatives
in the representation used. The more complicated case of an operator
$X$ with arbitrarily nonvanishing off-diagonal elements will be
discussed at the end of this section.

An important special case, relevant for example to electronic
structure calculations, is to choose for the operator $X$ the
Hamiltonian $\Ham$ and for $\S$ the $3N$-dimensional real-space
configuration of the system.  Then, the quantity $\XT$ is called the
{\em local energy,} denoted by $\Eloc$.  Clearly, in the ideal case
that $\ket{\uT}$ is an exact eigenvector of the evolution operator
$G$, and if $X$ commutes with $G$ then the configurational eigenvalue
$\XT(\S)$ is a constant independent of $\S$ and equals the true
eigenvalue of $X$.  In this case the variance of the Monte Carlo
estimator in Eq.~(\ref{eq.X.MCav}) goes to zero, which is an important
zero-variance principle satisfied by variational Monte Carlo.  The
practical implication is that the efficiency of the Monte Carlo
computation of the energy can be improved arbitrarily by improving the
quality of the trial function.  Of course, usually the time required
for the computation of $\uT(\S)$ increases as the approximation
becomes more sophisticated. For the energy the optimal choice
minimizes the product of variance and time; no such optimum exists for
an operator that does not commute with $G$ or if one makes the fixed
node approximation, described in Section~\ref{sec.fixednode}, since in
these cases the results have a systematic error that depends on the
quality of the trial wavefunction.

\subsection{Metropolis Method}
\label{sec.Metropolis}
A Monte Carlo process sampling the probability distribution
$\uT(\S)^2$ is usually generated by
means of the generalized Metropolis algorithm, as follows.  Suppose a
configuration $\S$ is given at time $t$ of the Monte Carlo process.
A new configuration $\Sp$ at time $t+1$ is generated by means of a
stochastic process that consists of two steps: (1) an intermediate
configuration $\Spp$ is proposed with probability $\Pi(\Spp|\S)$;
(2.a) $\Sp=\Spp$ with
probability $p\equiv A(\Spp|\S)$, {\it i.e.,} the proposed configuration is
accepted; (2.b) $\Sp=\S$ with probability $q\equiv 1-A(\Spp|\S)$, {\it
i.e.,}  the proposed configuration is rejected and the old
configuration $\S$ is promoted to time $t+1$.  More explicitly, the
Monte Carlo sample is generated by means of a Markov matrix $P$ with
elements $P(\Sp|\S)$ of the form
\begin{eqnarray}
P(\Sp|\S) = \left\{ \begin{array}{ll}
A(\Sp|\S) \; \Pi(\Sp|\S)  &{\rm for} \;\; \Sp \ne \S \\
1 \; - \; \sum_{\Spp \ne \S}
A(\Spp|\S) \; \Pi(\Spp|\S) &{\rm for} \;\; \Sp = \S,
\label{Markov.matrix}
\end{array}
\right.
\label{eq.P=AT}
\end{eqnarray}
if the states are discrete and
\begin{eqnarray}
P(\Sp|\S) = A(\Sp|\S) \; \Pi(\Sp|\S) \;+\;
\left[1 \; - \; \int d\Spp  A(\Spp|\S) \; \Pi(\Spp|\S)\right] \;
\delta(\Sp-\S)
\label{Markov.matrixc}
\end{eqnarray}
if the states are continuous.  Correspondingly,
in Eq.~(\ref{Markov.matrix}) $\Pi$ and $P$ are probabilities
while in Eq.~(\ref{Markov.matrixc}) they are probability densities.

The Markov matrix $P$ is designed to satisfy detailed balance
\beq
P(\Sp|\S) \; \uT(\S)^2=P(\S|\Sp) \; \uT(\Sp)^2,
\label{eq.det.balance.uT}
\eeq
so that, if the process has a unique stationary distribution, this
will be $\uT(\S)^2$, as desired.
In principle, one has great freedom in the choice of the proposal
matrix $\Pi$, but it is necessary to satisfy the requirement that
transitions can be made between (almost) any pair of states
with nonvanishing probability (density) in a finite number of steps.

Once a proposal matrix $\Pi$ is selected, an acceptance matrix is
defined so that detailed balance, Eq.~(\ref{eq.det.balance.uT}), is
satisfied
\beq
A(\Sp|\S)=
\min\left[1, \; {\Pi(\S|\Sp) \; \uT(\Sp)^2 \over \Pi(\Sp|\S) \; \uT(\S)^2}\right].
\label{eq.accept}
\eeq
For a given choice of $\Pi$, infinitely many choices can be made for
$A$ that satisfy detailed balance but the preceding choice is the one with
the largest acceptance.  We note that if the preceding algorithm is used,
then $\XT(\S_t)$ in the sum in Eq.~(\ref{eq.X.MCav}), can be replaced
by the expectation value conditional on $\Sppt$ having been proposed:
\beq
X_{\TT}^{(0,0)}=\lim_{L\to\infty}{1\over L}
\sum_{t=1}^L \left[p_t \XT(\Sppt)+q_t \XT(\S_t)\right],
\label{eq.X.MCav2}
\eeq
where $p_t=1-q_t$ is the probability of accepting $\Sppt$.
This has the advantage of reducing the statistical error somewhat, since
now $\XT(\Sppt)$ contributes to the average even for rejected moves,
and will increase efficiency if $\XT(\Sppt)$ is readily available.

If the proposal matrix $\Pi$ is symmetric, as is the case if one
samples from a distribution uniform over a cube centered at $\S$,
such as in the original Metropolis method~\cite{Metropolis}, the
factors of $\Pi$ in the numerator and denominator of
Eq.~(\ref{eq.accept}) cancel.

Finally we note that it is not necessary to sample the
distribution $u_{\T}^2$ to compute $X_{\TT}$: any distribution that
has sufficient overlap with $u_{\T}^2$ will do.  To make this point
more explicitly, let us introduce the average of some stochastic
variable $X$ with respect to an arbitrary distribution $\rho$:
\beq
\langle X \rangle_\rho \equiv
{\sum_{\S} X(\S) \rho(\S) \over \sum_{\S} \rho(\S)}
\eeq
The following relation shows that the desired results can be obtained
by reweighting, {\it i.e.}, any distribution $\rho$ will suffice as
long as the ratio $\uT(\S)^2/\rho(\S)$ does not fluctuate too wildly:
\beq
X_{\TT}^{(0,0)}=\langle X \rangle_{u_{\T}^2} =
{\langle X u_{\T}^2/\rho \rangle_\rho \over
\langle u_{\T}^2/\rho \rangle_\rho}.
\label{eq.reweight}
\eeq
This is particularly useful for calculation of the difference of
expectation values with respect to two closely related distributions.
An example of this~\cite{ReynoldsBarnettHammondLester86,Umrigar89}
is the calculation of the energy
of a molecule as a function of the inter-nuclear distance.

\subsection{Projector Monte Carlo and Importance Sampling}
\label{sec.projMC}

The generalized Metropolis method is a very powerful way to sample an
arbitrary {\it given} distribution, and it allows one to construct
infinitely many Markov processes with the desired distribution as the
stationary state.  None of these, however, may be appropriate to
design a \MC\ version of the power method to solve eigenvalue
problems.  In this case, the evolution operator $G$ is {\it given,}
possibly in approximate form, and its dominant eigenstate may {\it
not} be known.  To construct an appropriate \MC\ process, the first
problem is that $G$ itself is not a Markov matrix, \ie, it may violate
one or both of the properties $G(\Sp|\S) \ge 0$ and $\sum_\Sp G(\Sp|\S) = 1$.
This problem can be solved if we can find a factorization of the evolution
matrix $G$ into a Markov matrix $P$ and a weight matrix $g$ with
non-negative elements such that
\beq
G(\Sp|\S) = g(\Sp|\S) P(\Sp|\S).
\label{eq.gP}
\eeq
The weights $g$ must be finite, and this almost
always precludes use of the Metropolis method for continuous systems,
as can be understood as follows. Since
there is a finite probability that a proposed state will be
rejected, the Markov matrix $P(\Sp|\S)$ will contain
terms involving $\delta(\S-\Sp)$, but generically, $G$ will not have
the same structure and will not allow the definition of finite weights
$g$ according to Eq.~(\ref{eq.gP}).  However, the Metropolis algorithm
can be incorporated as a component of an algorithm if
an approximate stationary state is known
and if further approximations are made, as in the diffusion Monte
Carlo algorithm discussed in Section~\ref{sec.Improved}.

As a comment on the side we note that violation of the condition that
the weight $g$ be positive results in the notorious {\it sign
problem} in one of its guises, which is in most cases unavoidable in the treatment of
fermionic or frustrated systems.  Many ingenious
attempts~\cite{FermionSign} have been made to solve this problem, but
this is still a topic of active research.  However, as mentioned, we
restrict ourselves in this chapter to the case of evolution operators
$G$ with nonnegative matrix elements only.

We resume our discussion of the factorization given in
Eq.~(\ref{eq.gP}).  Suppose for the sake of argument that the left
eigenstate $\hat{\psi_0}$ of $G$ is known and that its elements are
positive,
\beq
\sum_{\Sp} \hat{\psi_0}(\Sp) G(\Sp|\S) = \lambda_0 \hat{\psi_0}(\S).
\eeq
If in addition, the matrix elements of $G$ are nonnegative,
the following matrix $\Phat$ is a Markov matrix
\beq
\Phat(\Sp|\S)={1\over \lambda_0}
\hat{\psi_0}(\Sp) G(\Sp|\S) {1\over \hat{\psi_0}(\S)}.
\eeq
Unless one is dealing with a Markov matrix from the outset, the left
eigenvector of $G$ is seldom known, but it is convenient, in any
event, to perform a so-called {\it importance sampling} transformation
on $G$.  For this purpose we introduce a guiding function $\ug$ and
define
\beq
\Ghat(\Sp|\S)= \ug(\Sp) G(\Sp|\S) {1\over \ug(\S)}.
\label{eq.Ghatg}
\eeq
We shall return to the issue of the guiding function, but for the time
being the reader can think of it either as an arbitrary, positive
function, or as an approximation to the dominant eigenstate of $G$.
{}From a mathematical point of view, anything that can be computed
with the original \MC\ evolution operator $G$ can also be computed
with $\Ghat$, since the two represent the same abstract operator in a
different basis. The representations differ only by
normalization constants.  All we have to do is to write all
expressions derived above in terms of this new basis.

We continue our discussion in terms of the transform $\Ghat$ and
replace Eq.~(\ref{eq.gP}) by the factorization
\beq
\Ghat(\Sp|\S) = \ghat(\Sp|\S) \Phat(\Sp|\S)
\label{eq.ghatPhat}
\eeq
and we assume that $\Phat$ has the explicitly known distribution
$\ug^2$ as its stationary state. The guiding function $\ug$ appears in
those expressions, and it should be kept in mind that they can be
reformulated by means of the reweighting procedure given in
Eq.~(\ref{eq.reweight}) to apply to processes with different
explicitly known stationary states.  On the other hand, one might be
interested in the infinite projection limit $p\to \infty$.  In that
case, one might use a \MC\ process for which the stationary
distribution is not known explicitly.  Then, the expressions below
should be rewritten so that the unknown distribution does not
appear in expressions for the time averages.
The function $\ug$ will still appear, but only as a transformation
known in closed form and no longer as the stationary state of $P$.
Clearly, a process for
which the distribution is not known in closed form cannot be used to
compute the matrix elements $\X \alpha {\beta}{p'}p$ for finite $p$
and $p'$ and given states $\ket{\ua}$ and $\ket{\ub}$.

One possible choice for $\Phat$ that avoids the Metropolis method and
produces finite weights is the following {\it generalized heat bath}
transition matrix
\beq
\Phat(\Sp|\S)=
{\Ghat(\Sp|\S) \over \sum_{\S_1} \Ghat(\S_1|\S)}.
\label{eq.PuG}
\eeq
If $G(\Sp|\S)$ is symmetric, this transition matrix has a known
stationary distribution, {\it viz.,} $\Gg(\S) \ug^2(\S)$, where
$\Gg(\S)=\bra{\S}| G \ket{\ug}/\ug(\S)$, the configurational
eigenvalue of $G$ in state $\S$.  $\Phat$ must be
chosen such that the corresponding transitions can be sampled
directly.  This is usually not feasible unless $\Phat$ is sparse or
near-diagonal, or can be transformed into a form involving
non-interacting degrees of freedom.  We note that if $\Phat$ is
defined by Eq.~(\ref{eq.PuG}), the weight matrix $\ghat$ depends only
on $\S$.

\subsection{Matrix Elements}

We now address the issue of computing the matrix elements $\X \alpha
\beta {p'} p$, assuming that the stationary state $\ug^2$ is known
explicitly and that the weight matrix $\ghat$ has finite elements.  We
shall discuss the following increasingly complex possibilities: (a)
$[X,G]=0$ and $X$ is near-diagonal in the $\S$ representation; (b) $X$
is diagonal in the $\S$ representation; (c) $X(\S|\Sp)$ is truly
off-diagonal.  The fourth case, {\it viz.}, $[X,G]=0$ and $X$ is not
near-diagonal is omitted since it can easily be constructed from the
three cases discussed explicitly.  When discussing case (c) we shall
introduce the concept of {\it side walks} and explain how these can be used to
compute matrix elements of a more general nature than discussed up to that point.
After deriving the expressions, we shall discuss the practical
problems they give rise to, and ways to reduce the variance of the
statistical estimators.  Since this yields expressions in a more
symmetric form, we introduce the transform
\beq
\Xhat(\Sp|\S)= \ug(\Sp) X(\Sp|\S) {1\over \ug(\S)}.
\eeq

\subsubsection{$[X,G]=0$ and $X$ Near-Diagonal}
\label{section.commute}
In this case, $\X \alpha {\beta}0 {p'+p} = \X \alpha \beta{p'} p $ and
it suffices to consider the computation of $\X \alpha \beta0 p $.  By
repeated insertion in Eq.~(\ref{eq.general.estimator}) of the
resolution of the identity in the $\S$-basis, one obtains the
expression
\beq
\X \alpha \beta 0 p =
{\sum_{\S_{p},\dots,\S_{0}}
\ua(\S_p) \;
X_\alpha (\S_{p}) \;
[\prod_{i=0}^{p-1} G(\S_{i+1}|\S_i)] \;
\ub(\S_0)
\over
{\sum_{\S_{p},\dots,\S_{0}}
\ua(\S_p) \;
[\prod_{i=0}^{p-1} G(\S_{i+1}|\S_i)] \;
\ub(\S_0)}}
\label{eq.XT0p.Srep}
\eeq
In the steady state, a series of subsequent states
$\S_t,\S_{t+1},\dots,\S_{t+p}$ occurs with probability
\beq
\mbox{Prob}(\S_t,\S_{t+1},\dots,\S_{t+p}) \propto
[\prod_{i=0}^{p-1} \Phat(\S_{t+i+1}|\S_{t+i})] \; \ug(\S_{t})^2.
\label{eq.prob.sequence}
\eeq
To relate products of the matrix $P$ to those of $G$, it is
convenient to introduce the following definitions
\beq
\W_t(p,q)=\prod_{i=q}^{p-1} \ghat(\S_{t+i+1}|\S_{t+i})
\eeq
Also, we define
\beq
\uhat_\omega(\S)={u_\omega(\S) \over \ug(\S)},
\eeq
where $\omega$ can be any of a number of subscripts.

With these definitions, combining Eqs.~(\ref{eq.gP}),
(\ref{eq.XT0p.Srep}), and (\ref{eq.prob.sequence}), one finds
\beq
X_{\alpha\beta}^{(0,p)}=
\lim_{L\to\infty}
{\sum_{t=1}^L \;
\uhata(\S_{t+p}) \;
X_\alpha(\S_{t+p}) \;
\W_t(p,0)
\uhatb(\S_{t})
\over
\sum_{t=1}^L \;
\uhata(\S_{t+p}) \;
\W_t(p,0)
\uhatb(\S_{t})}.
\label{eq.XT0p.MC}
\eeq

\subsubsection{Diagonal $X$}
\label{section.diag}
The preceding discussion can be generalized straightforwardly to the case
in which $X$ is diagonal in the $\S$ representation.  Again by
repeated insertion of the resolution of the identity in the $\S$-basis
in the Eq.~(\ref{eq.general.estimator}) for $\X \alpha \beta{p'} p$,
one obtains the identity
\beq
X_{\alpha\beta}^{(p',p)}=
{\sum_{\S_{p'+p},\dots,\S_{0}} \;
\ua(\S_{p'+p}) \;
[\prod_{i=p}^{p'+p-1} G(\S_{i+1}|\S_i)] \;
X(\S_{p}|\S_{p}) \;
[\prod_{i=0}^{p-1} G(\S_{i+1}|\S_i)] \;
\ub(\S_0)
\over
\sum_{\S_{p'+p},\dots,\S_{0}} \;
\ua(\S_{p'+p}) \;
[\prod_{i=0}^{p'+p-1} G(\S_{i+1}|\S_i)] \;
\ub(\S_0)}.
\label{eq.XTTpp.Srep}
\eeq
Again by virtue of Eq.~(\ref{eq.prob.sequence}), we find
\beq
X_{\alpha\beta}^{(p',p)}=
\lim_{L\to\infty}
{\sum_{t=1}^L \;
\uhata(\S_{t+p'+p}) \;
\W_t(p'+p,p)
X(\S_{t+p}|\S_{t+p}) \;
\W_t(p,0)
\uhatb(\S_t)
\over
\sum_{t=1}^L
\uhata(\S_{t+p'+p}) \;
\W_t(p'+p,0)
\uhatb(\S_t)}.
\label{eq.XTTpp.MC}
\eeq

\subsubsection{Nondiagonal $X$}
\label{section.non-diag}
If the matrix elements of $G$ vanish only if those of $X$ do,
the preceding method can be generalized immediately to the final case in
which $X$ is nondiagonal.  Then, the analog of
Eq.~(\ref{eq.XTTpp.MC}) is
\beq
X_{\alpha\beta}^{(p',p)}=
\lim_{L\to\infty}
{\sum_{t=1}^L \;
\uhata(\S_{t+p'+p}) \;
\W_t(p'+p,p+1)
x(\S_{t+p+1}|\S_{t+p}) \;
\W_t(p,0)
\uhatb(\S_t)
\over
\sum_{t=1}^L \;
\uhata(\S_{t+p'+p})\;
\W_t(p'+p,0)
\uhatb(\S_t)}.
\label{eq.XTTppn.MC}
\eeq
where the $x$ matrix elements are defined by
\beq
x(\Sp|\S) ={X(\Sp|\S) \over P(\Sp|\S)}= {\Xhat(\Sp|\S) \over \Phat(\Sp|\S)}.
\label{eq.xhat}
\eeq
Clearly, the preceding definition of $x(\Sp|\S)$ fails when
$\Phat(\Sp|\S)$ vanishes but $\Xhat(\Sp|\S)$ does not.  If that can
happen, a more complicated scheme can be employed in which one
introduces {\it side-walks}.  This is done by interrupting the
continuing stochastic process at time $t+p$ by introducing a finite
series of auxiliary states $\S'_{t+p+1},\dots,\S'_{t+p'+p}$.  The
latter are generated by a separate stochastic process so that in
equilibrium, the sequence of subsequent states
$\S_t,\S_{t+1},\dots,\S_{t+p},\S'_{t+p+1},\dots,\S'_{t+p'+p}$ occurs
with probability
\begin{eqnarray}
&\mbox{Prob}[(\S_t,\S_{t+1},\dots,\S_{t+p},\S'_{t+p+1},
\dots,\S'_{t+p'+p})]\propto \nonumber\\
&
[\prod_{i=p+1}^{p'+p-1}\Phat(\S'_{t+i+1}|\S'_{t+i})]\;
\Phat_X(\S'_{t+p+1}|\S_{t+p})\;
[\prod_{i=0}^{p-1}\Phat(\S_{t+i+1}|\S_{t+i})]
\ug(\S_t)^2
\label{eq.prob'.sequence}
\end{eqnarray}
where $\Phat_X$ is a Markov matrix chosen to replace $\Phat$ in
Eq.~(\ref{eq.xhat}) so as to yield finite weights $x$.  In this
scheme, one generates a continuing thread identical to the usual \MC\
process in which each state $\S_t$ is sampled from the stationary
state of $\Phat$, at least if one ignores the initial equilibration.  Each
state $\S_t$ of this backbone forms the beginning of a side walk, the
first step of which is sampled from $\Phat_X$, while $\Phat$ again generates
subsequent ones.  Clearly, with respect to the side walk, the first
step disrupts the stationary state, so that the $p'$ states
$\S'_{t'}$, which form the side walk, do not sample the stationary
state of the original stochastic process generated by $\Phat$, unless
$\Phat_X$ coincidentally has the same stationary state as $\Phat$.

A problem with the matrix elements we dealt with up to now is that in
the limit $p'$ or $p \to \infty$ all of them reduce to matrix elements
involving the dominant eigenstate, although symmetries might be used
to yield other eigenstates besides the absolute dominant one.
However, if symmetries fail, one has to employ the equivalent of an
orthogonalization scheme, such as, discussed in the next section,
or one is forced to resort to evolution operators that contain, in
exact or in approximate form, the corresponding projections.  An
example of this are matrix elements computed in the context of the
fixed-node approximation\cite{ReynoldMatrixElements}, discussed in
Section~\ref{sec.fixednode}.  Within the framework of this
approximation, one considers quantities of the form
\beq
X^{(p',p)}_{\alpha \beta} =
{\bra{\ua}|G_1^{p'}XG_2^{p}\ket{\ub} \over
\sqrt{\bra{\ua}|G_1^{2p'}\ket{\ua} \bra{\ub} |G_2^{2p}\ket{\ub}}},
\label{eq.general.estimator2}
\eeq
where the $G_i$ are evolution operators combined with appropriate
projectors, which in the fixed-node approximation are defined by the
nodes of the states $\ua(\S)$ and $\ub(\S)$.
We shall describe how the preceding expression,
Eq.~(\ref{eq.general.estimator2}), can be evaluated, but rather than
writing out all the expressions explicitly, we present just the
essence of the \MC\ method.

To deal with these expressions, one
generates a backbone time series of states sampled from any
distribution, say, $\ug(\S)^2$, that has considerable overlap with the
the states $|\ua(\S)|$ and $|\ub(\S)|$.  Let us distinguish those backbone
states by a superscript $0$. Consider any such state $S^{(0)}$
at some given time.  It forms the starting point of two side walks.
We denote the states of these side walks by $\S_{t_i}^{(i)}$
where $i=1,2$ identifies the side walk and $t_i$ labels the side steps.
The side walks are generated from factorizations of the usual form,
defined in Eq.~(\ref{eq.ghatPhat}), say $\Ghat_i=\ghat_i \Phat_i$.  A
walk
\beq
{\cal S}=
[\S^{(0)},(\S_1^{(1)},\S_2^{(1)},\dots),
(\S_1^{(2)},\S_2^{(2)},\dots)]
\eeq
occurs with probability
\beq
\mbox{Prob}({\cal S})=
\ug(\S{(0)})^2 \;
\Phat_1(\S^{(0)}|\S_1^{(1)})\dots  \Phat_2(\S^{(0)}|\S_1^{(2)})\dots.
\eeq
We leave it to the reader to show that this probability
suffices for the computation of all expressions appearing in numerator
and denominator of Eq.~(\ref{eq.general.estimator2}), in the case that
$X$ is diagonal, and to generate the appropriate generalizations to
other cases.

In the expressions derived above, the power method projections
precipitate products of reweighting factors $\ghat$, and, as the
projection times $p$ and $p'$ increase, the variance of the Monte
Carlo estimators grows at least exponentially in the square root of the
projection time.  Clearly, the presence of the
fluctuating weights $\ghat$ is due to the fact that the evolution
operator $\Ghat$ is not Markovian in the sense that it fails to
conserve probability. The importance sampling transformation
Eq.~(\ref{eq.Ghatg}) was introduced to mitigate this problem.
In Section~\ref{sec.branching}, an algorithm involving branching walks
will be introduced, which is a different
strategy devised to deal with this problem.  In diffusion and transfer
matrix Monte Carlo, both strategies, importance sampling and
branching, are usually employed simultaneously.

\subsection{Excited States}

Given a set of basis states, excited eigenstates can be computed
variationally by solving a linear variational problem and the
Metropolis method can be used to evaluate the required matrix
elements.  The methods involving the power method, as described above,
can then be used to remove the variational bias systematically.
\cite{CeperleyBernu88,BerCepLes90,BrGlLes95}

In this context matrix elements appear in the solution of the
following variational problem. As was mentioned several times before,
the price paid for reducing the variational bias is increased
statistical noise, a problem which appears in this context with a
vengeance.  Again, the way to keep this problem under control is the
use of optimized trial vectors.

The variational problem to be solved is the following one.  Given $n$
basis functions $\ket {u_i}$, find the $n\times n$ matrix of
coefficients $d_i^{(j)}$ such that
\beq
\ket{\tilde \psi_j} = \sum_{i=1}^n d_i^{(j)} \ket{ u_i}
\label{eq.varexc}
\eeq
are the best variational approximations for the $n$ lowest eigenstates
$\ket{\psi_i}$ of some Hamiltonian $\Ham$. In this problem we
shall use the language of the quantum mechanical systems, where
one has to distinguish the Hamiltonian from the evolution operator
$\exp(-\tau \Ham)$.  In the statistical mechanical applications, one
has only the equivalent of the latter.  In the expressions to be
derived below the substitution $\Ham G^p \to G^{p+1}$ will produce the
expressions required for the statistical mechanical applications, at
least if we assume that the nonsymmetric matrices that appear in that
context have been symmetrized.\footnote{This is not possible in general
for transfer matrices of
systems with helical boundary conditions, but the connection between
left and right eigenvectors of the transfer matrix (see
Section~\ref{sec.transfer}) can be used to generalize the approach
discussed here.}

One seeks a solution to the linear variational problem in
Eq.~(\ref{eq.varexc}) in the sense that for all $i$ the Rayleigh
quotient $\bra {\tilde \psi_{i}}|\Ham \ket {\tilde \psi_i}/
\bra {\tilde \psi_{i}}\ket {\tilde \psi_i}$ is
stationary with respect to variation of the coefficients $d$.  The
solution is that the matrix of coefficients $d$ has to satisfy the
following generalized eigenvalue equation
\beq
\sum_{i=1}^n H_{ki} d_i^{(j)} = {\tilde E}_j
\sum_{i=1}^n N_{ki}  d_i^{(j)},
\label{eq.geneig}
\eeq
where
\beq
H_{ki}=\bra {u_k}|\Ham \ket {u_i},
\eeq
and
\beq
N_{ki}=\bra {u_k}\ket {u_i}.
\eeq

Before discussing \MC\ issues, we note a number of important
properties of this scheme.  First of all, the basis states $\ket
{u_{i}}$ in general are not orthonormal and this is reflected by the
fact that the matrix elements of $N$ have to be computed.  Secondly,
it is clear that any nonsingular linear combination of the basis
vectors will produce precisely the same results, obtained from the
correspondingly transformed version of Eq.~(\ref{eq.geneig}).  The
final comment is that the variational eigenvalues bound the exact
eigenvalues, {\it i.e.,} $\tilde E_i \ge E_i$.  One recovers exact
eigenvalues $E_i$ and the corresponding eigenstates, if the
$\ket{u_{i}}$ span the same space as the exact eigenstates.

The required matrix elements can be computed using the variational
\MC\ method
discussed in the previous section.  Furthermore, the power method can
be used to reduce the variational bias.  Formally, one simply defines
new basis states
\beq
\ket {u_{i}^{(p)}} = G^p \ket {u_{i}}
\eeq
and substitutes these new basis states for the original ones.  The
corresponding matrices
\beq
H_{ki}^{(p)}=\bra {u_k^{(p)}}|\Ham \ket {u_i^{(p)}}
\eeq
and
\beq
N_{ki}^{(p)}=\bra {u_k^{(p)}}\ket {{u_i}^{(p)}}
\eeq
can again be computed by applying the methods introduced in
Section~\ref{sec.singlethread} for the computation of general matrix
elements by a \MC\ implementation of the power method.

As an explicit example illustrating the nature of the \MC\
time-averages that one has to evaluate in this approach, we write down
the expression for $N_{ij}^{(p)}$ as used for the computation of
eigenvalues of the Markov matrix relevant to the problem of critical
slowing down:
\beq
N_{ij}^{(p)} \approx \sum_t {u_i(\S_t) \over \psiB(\S_t)}
{u_j(\S_{t+p}) \over \psiB (\S_{t+p})},
\label{eq.uucorrel}
\eeq
where the $\S_t$ are configurations forming a time series which, as we
recall, is designed to sample the distribution of a system in
thermodynamic equilibrium, {\it i.e.,} the Boltzmann distribution
$\psiB^2$.  The expression given in Eq.~(\ref{eq.uucorrel}) yields the
$u/\psiB$-auto-correlation function at lag $p$.  The expression for
$H_{ij}^{(p)}$ is similar, and represents a cross-correlation function
involving the configurational eigenvalues of the Markov matrix in the
various basis states.  Compared to the expressions derived in
Section~\ref{sec.singlethread}, Eq.~(\ref{eq.uucorrel}) takes a
particularly simple form in which products of fluctuating weights are
absent, because in this particular problem one is dealing with a
probability conserving evolution operator from the outset.

Eq.~(\ref{eq.uucorrel}) shows why this method is sometimes called
correlation function \MC, but it also illustrates
a new feature, namely, that it is efficient to
compute all required matrix elements simultaneously.  This can be done
by generating a Monte Carlo process with a known distribution
which has sufficient overlap with all $|u_i(\S)|\equiv |\bra {\S}
\ket{u_i}|$.  This can be arranged, for example, by sampling a guiding
function $u_{\rm g}(\S)$ defined by
\beq
u_{\rm g}(\S) = \sqrt{\sum_{i=1}^n a_i u_i(\S)^2},
\eeq
where the coefficients $a_i$ approximately normalize the
basis states $\ket{u_i}$, which may require a preliminary \MC\ run.
See Ceperley and
Bernu \cite{CeperleyBernu88} for an alternative choice for a guiding
function.  In the computations \cite{NB97.prl}
to obtain the spectrum of the Markov
matrix in critical dynamics, as illustrated by Eq.~(\ref{eq.uucorrel}),
the Boltzmann distribution, is used as a guiding function.  It apparently
has enough overlap with the decaying modes that no special purpose
distribution has to be generated.

\subsection{How to Avoid Reweighting}

Before discussing the branching algorithms designed to deal more
efficiently with the reweighting factors appearing in the expressions
discused above, we briefly mention an alternative that has surfaced
occasionally without being studied extensively, to our knowledge.  The
idea will be illustrated in the case of the computation of the matrix
element $\X \alpha \beta 0 p$, and we take Eqn.~(\ref{eq.XT0p.Srep})
as our starting point.  In statistical mechanical language, we
introduce a reduced Hamiltonian
\beq
\H=\ln \ug(\S_p)+\sum_{i=0}^{p-1} \ln G(\S_{i+1}|\S_i)+\ln \ug(\S_0)
\eeq
and the corresponding Boltzmann distribution
$\exp{-\H(\S_p,\dots,\S_0)}$.  One can now use the standard Metropolis
algorithm to sample this distribution for this system consisting of
$p+1$ layers bounded by the layers $0$ and $p$.  For the evaluation
of Eq.~(\ref{eq.XT0p.Srep}) by Monte Carlo, this expression then
straightforwardly becomes a ratio of correlation functions involving
quantities defined at the boundaries.  To see this, all one has to do
is to divide the numerator and denominator of Eq.~(\ref{eq.XT0p.Srep})
by the partition function
\beq
Z=\sum_{\S_p,\dots,\S_0} e^{-\H(\S_p,\dots,\S_0)}
\eeq
Note that in general, boundary terms involving some appropriately
defined $\ug$ should be introduced to ensure the non-negativity of the
distribution.  For the simultaneous computation of matrix elements for
several values of the indices $\alpha$ and $\beta$, a guiding function
$\ug$ should be chosen that has considerable overlap with the
corresponding $|\ua|$ and $|\ub|$.

The Metropolis algorithm can of course be used to sample any
probability distribution, and the introduction of the previous
Hamiltonian illustrates just one particular point of view.
If one applies the preceding idea to the case of the imaginary-time
quantum mechanical evolution operator, one obtains a modified version of
the standard path-integral Monte Carlo method, in which case the layers
are usually called time slices.  Clearly, this method has the
advantage of suppressing the fluctuating weights in estimators.
However, the disadvantage is that sampling the full, layered system
yields a longer correlation time than sampling the single-layer
distribution $\ug^2$.
This is a consequence of the fact that the microscopic
degrees of freedom are more strongly correlated in a layered system
than in a single layer.  Our limited experience suggests that for
small systems reweighting is more efficient, whereas the Metropolis
approach tends to become more efficient as the system grows in size.
\cite{NOY}

\section{Trial Function Optimization}
\label{sec.optimiz}
In the previous section it was shown that eigenvalue estimates can be
obtained as the eigenvalues of the matrix ${N^{(p)}}^{-1} H^{(p)}$.
The variational error in these estimates decreases as $p$ increases.
In general, these expressions involve weight products of increasing
length, and consequently the errors grow exponentially, but even in
the simple case of a probability conserving evolution operator, errors
grow exponentially.  This is a consequence of the fact that the
auto-correlation functions in $N^{(p)}$, and the cross-correlation functions
in $H^{(p)}$, in the limit $p\to \infty$
reduce to quantities that contain an exponentially vanishing amount of
information about the
subdominant or excited-state eigenvalues, since the spectral weight of
all but the dominant eigenstate is reduced to zero by the power method.

The practical implication is
that this information has to be retrieved with sufficient accuracy
for small values of $p$, before the signal disappears in
the statistical noise.  The projection time $p$ can be kept small by
using optimized basis states constructed to reduce the overlap of
the linear space spanned by the basis states $\ket{u_i}$ with the space
spanned by the eigenstates beyond the first $n$ of interest.  We shall
describe, mostly qualitatively, how this can be done by a generalization of
a method used for optimization of individual basis
states\cite{NU96,Coldwell,UWW,N96}, {\it viz.} minimization of
variance of the configurational eigenvalue, the local energy in
quantum Hamiltonian problems.

Suppose that $\uT(\S,v)$ is the value of the trial function $\uT$ for
configuration $\S$ and some choice of the parameters $v$ to be
optimized. As in Eq.~(\ref{eq.XT}), the {\it configurational eigenvalue}
$\lambda(S,v)$ of configuration $\S$ is defined by
\beq
\uT'(\S,v) \equiv \lambda(\S,v) \uT(\S,v),
\label{eq.lamdas}
\eeq
where a prime is used to denote, for arbitrary $\ket{f}$, the components
of $G\ket f$, or $\Ham\ket f$ as is more convenient for quantum
mechanical applications.
The optimal values of the variational parameters are
obtained by minimization of the variance of $\lambda(\S,v)$, estimated
as an average over a small Monte Carlo sample.  In the ideal case,
{\it i.e.,} if an exact eigenstate can be reproduced by some choice of
the parameters of $\uT$, the minimum of the variance yields the exact
eigenstate not only if it were to be computed exactly, but even if it is
approximated by summation over a Monte Carlo sample.  A similar
zero-variance principle holds for the method of simultaneous
optimization of several trial states to be discussed next.  This is in
sharp contrast with the more traditional Rayleigh-Ritz extremization
of the Rayleigh quotient, which frequently can produce arbitrarily
poor results if minimized over a small sample of configurations.

For conceptual simplicity, we first generalize the preceding method to the
more general ideal case that reproduces the exact values of the
desired $n$ eigenstates of the evolution operator $G$.  As a
byproduct, our discussion will produce an alternative to the
derivation of Eq.~(\ref{eq.geneig}).  To compute $n$ eigenvalues, we
have to optimize the $n$ basis states $\ket{u_i}$, where we have
dropped the index ``T'', and again we assume we have a sample of
$M$ configurations $\S_\alpha$, $\alpha=1,\dots,M$
sampled from $\ug^2$.
The case we consider is ideal in the sense that we assume that these
basis states $\ket{u_i}$ span an $n$-dimensional invariant subspace of
$G$.  In that case, by definition there exists a matrix $\Lambda$ of
order $n$ such that
\beq
{u_i}'(\S_\alpha)=\sum_{j=1}^n \Lambda_{ij} u_j(\S_\alpha).
\label{eq.Lam_matrix}
\eeq
Again, the prime on the left-hand side of this equation indicates
multiplication by $G$ or by $\Ham=-\tau^{-1}\ln G$.  If the number of
configurations is large enough, $\Lambda$ is
for all practical purposes determined uniquely by the set of equations
(\ref{eq.Lam_matrix}) and one finds
\beq
\Lambda= N^{-1} H,
\label{eq.Lam=H/N}
\eeq
where
\begin{eqnarray}
& N_{ij}={1 \over M}\sum_{\alpha=1}^M u_i(S_\alpha) u_j(S_\alpha)/
\ug(S_\alpha)^2,
\nonumber \\
& H_{ij}={1 \over M}\sum_{\alpha=1}^M u_i(S_\alpha) u_j '(S_\alpha)/
\ug(S_\alpha)^2.
\label{eq.NP}
\end{eqnarray}
In the limit $M\to \infty$ this indeed reproduces the matrices $N$ and
$H$ in Eq.~(\ref{eq.geneig}).  In the nonideal case, the
space spanned by the $n$ basis states $\ket{u_i}$ is not an invariant
subspace of the matrix $G$.  In that case, even though
Eq.~(\ref{eq.Lam_matrix}) generically has no true solution,
Eqs.~(\ref{eq.Lam=H/N}) and (\ref{eq.NP}) still constitute a solution
in the least-squares sense, as the reader is invited to show
for himself by solving the normal equations.

Next, let us consider the construction of a generalized optimization
criterion.  As mentioned before, if a set of basis states span an
invariant subspace, so does any nonsingular linear combination.  In
principle, the optimization criterion should have the same invariance.
The {\it matrix} $\Lambda$ lacks this property, but its {\it spectrum}
is invariant. Another consideration is that, while the local
eigenvalue is defined by a single configuration $\S$, it takes at least $n$
configurations to determine the ``local'' matrix $\Lambda$.  This
suggests that one subdivide the sample into subsamples of at least $n$
configurations each and minimize the variance of the {\it local
spectrum} over these {\it subsamples.}  Again in principle, this has
the advantage that the optimization can exploit the fact that linear
combinations of the basis states have more variational freedom to
represent the eigenstates than does each variational basis function
separately.  In practice, however, this advantage seems to be negated
by the difficulty of finding good optimal parameters. This is a
consequence of the fact that invariance under linear transformation
usually can be mimicked by variation of the parameters of the basis
states.  In other words, a linear combination of basis states can be
represented accurately, at least relative to the noise in the local
spectrum, by a single basis state with appropriately chosen
parameters. Consequently, intrinsic flaws of the trial states exceed
what can be gained in reducing the variance of the local spectrum by
exploiting the linear variational freedom, most of which is already
used anyway in the linear variational problem that was discussed at
the beginning of this section. This means that one has to contend with
a near-singular nonlinear optimization problem.  In practice, to
avoid the concomitant slow convergence, it seems to be more efficient
to break the ``gauge symmetry'' and select a preferred basis, which
most naturally is done by requiring that each basis state {\em itself}
is a good approximate eigenstate.

The preceding considerations, of course, leave us with two criteria, {\it
viz.}  minimization of the variance of the local spectrum as a whole,
and minimization of the variance of the configurational eigenvalue
separately. To be of practical use, both criteria have to be combined,
since if one were to proceed just by minimization of the variance of
the configurational eigenvalues separately, one would simply keep
reproducing the same eigenstate.  In a non\MC\ context this can be
solved simply by some orthogonalization scheme, but as far as \MC\ is
concerned, that is undesirable since it fails to yield a
zero-variance optimization principle.

\section{Branching Monte Carlo}
\label{sec.branching}

In Section~\ref{sec.singlethread} we discussed a method to
compute Monte Carlo averages by exploiting
the power method to reduce the spectral weight of undesirable,
subdominant eigenstates.  We saw that this leads to products of weights
of subsequent configurations sampled by a \MC\ time series.
To suppress completely
the systematic errors due to finite projection times, {\it i.e.,} the
variational bias, one has to take averages of infinite products of
weights.  This limit would produce an ``exact'' method with infinite
variance, which is of no practical use.

We have also discussed how optimized trial states can be used to
reduce the variance of this method.  The variance reduction may come
about in two ways.  In the first place, by starting with optimized
trial states of higher quality, the variational bias is smaller to
begin with so that fewer power method projections are required.
In practical terms, this leads to a reduction of the
number of factors in the fluctuating products.  Secondly, a good
estimate of the dominant eigenstate, can be used to reduce the amount
by which the evolution operator, divided by an appropriate constant,
violates conservation of probability, which reduces the variance of
the individual fluctuating weight factors.  All these considerations
also apply to the branching Monte Carlo algorithm discussed in this
section, which can be modified accordingly and in complete analogy
with our previous discussion.

Before discussing the details of the branching algorithm, we mention
that the algorithm presented here\cite{NB96.prb} contains the
mathematical essence of both the diffusion and transfer matrix Monte
Carlo algorithms.  A related algorithm, {\it viz.}, Green function
Monte Carlo, adds yet another level of complexity due to the
fact that the evolution operator is known only
as an infinite series.  This series is stochastically summed at each
step of the power method iterations.  In practice this implies that
even the time step becomes stochastic and intermediate \MC\ configurations
are generated that do not contribute to expectation values.
Neither Green function Monte
Carlo, nor its generalization designed to compute quantities at
non-zero temperature \cite{WhitlockKalos79}, will be discussed in this
chapter and we refer the interested reader to the literature for further
details.\cite{CeperleyKalosbook,SchmidtMoskowitz,KLV74,Ceperley83}

Let us consider in detail the mechanism that produces large variance.
This will allow us to explain what branching accomplishes if one has
to compute products of many (ideally infinitely many) fluctuating
weights.  The time average over these products will typically be
dominated by only very few large terms; the small terms are equally
expensive to compute, but play no significant role in the average.
This problem can be solved by performing many simultaneous Monte Carlo
walks.
One evolves a collection of walkers from one
generation to the next and the key idea is to eliminate the
light-weight walkers which produce relatively small contributions to
the time average.  To keep the number of walkers reasonably constant,
heavy-weight walkers are duplicated and the clones are subsequently
evolved (almost) independently.

An algorithm designed according to this concept does not cut off the
products over weights and therefore seems to correspond to infinite
projection time.  It would therefore seem that the time average over a
stationary branching process corresponds to an average over the exact
dominant eigenstate of the \MC\ evolution operator, but, as we shall see,
this is rigorously the case only in the limit of an infinite number of
walkers\cite{Hetherington84,NBPRB33}; for any finite number of walkers,
the stationary distribution has a bias inversely proportional to
the number of walkers, the so-called {\em population
control bias.}  If the fluctuations in the weights are small and
correlations (discussed later) decay rapidly, this bias tends to be small.
In many
applications this appears to be the case and the corresponding bias is
statistically insignificant.  However, if these methods are applied to
statistical mechanical systems at the critical point, significant bias
can be introduced.  We shall discuss a simple method of nearly
vanishing computational cost to detect this bias and correct for it
in all but the worst-cases scenarios.

To discuss the branching Monte Carlo version of the power method, we
continue to use the notation introduced above and again consider the
\MC\ evolution operator $G(\Sp|\S)$.
As above, the states $\S$ and $\S'$ will be treated here as discrete,
but in practice the distinction between continuous and discrete states
is a minor technicality, and generalization to
the continuous case follows immediately by replacing sums by integrals
and by replacing Kronecker $\delta$'s by Dirac $\delta$ functions.

To implement the power method iterations in
Eq.~(\ref{eq.power.method}) by a branching Monte Carlo process,
$\ket{u^{(t)}}$ is represented by a collection of $N_t$ walkers, where
a walker by definition is a state-weight pair
$(\S_\alpha,w_\alpha),\,\alpha=1,\dots ,N_t$. As usual, the state
variable $\S_\alpha$ represents a possible configuration of the system
evolving according to $G$, and $w_\alpha$ represents the statistical
weight of walker $\alpha$. These weights appear in averages and the
efficiency of the branching Monte Carlo algorithm is realized by
maintaining the weights in some range $w_{\rm l}<w_\alpha<w_{\rm u}$,
where $w_{\rm l}$ and $ w_{\rm u}$ are bounds introduced so as to keep
all weights $w_\alpha$ of the same order of magnitude.

The first idea is to interpret a collection of walkers that make up
generation $t$ as a representation of the (sparse) vector
$\ket{\uunder^{(t)}}$ with components
\beq
\bra \S \ket{\uunder^{(t)}} \equiv
\underline u^{({t})}(\S) = \sum_{\alpha=1}^{N_t} w_\alpha
\delta_{\S,\S_\alpha},
\label{eq.stoch.vector}
\eeq
where $\delta$ is the usual Kronecker $\delta$-function.  The
underbar is used to indicate that the $\underline u^{({t})}(\S)$
represent a stochastic vector $\ket{\uunder^{(t)}}$.  Of course, the
same is true formally for the single thread \MC.  The new feature is
that one can think of the collective of walkers as a reasonably
accurate representation of the stationary state at each time step,
rather than in the long run.

The second idea is to define a stochastic process in which the walkers
evolve with transition probabilities such that the expectation value
of $c_{t+1}\ket{\uunder^{({t+1})}}$, as represented by the walkers of
generation $t+1$, equals $G \ket{\uunder ^{({t})}}$ for any given
collection of walkers representing $\ket{\uunder^{({t})}}$.  It is
tempting to conclude that, owing to this construction, the basic
recursion relation of the power method, Eq.~(\ref{eq.power.method}),
is satisfied in an average sense, but this conclusion is not quite
correct.  The reason is that in practice, the constants $c_t$ are
defined on the fly.  Consequently, $c_{t+1}$ and
$\ket{\uunder^{({t+1})}}$ are correlated random variables and
therefore there is no guarantee that the stationary state expectation
value of $|{\underline u}^{({t})}\rangle$ is {\em exactly} an
eigenstate of $G$, except in the limit of nonfluctuating
normalization constants $c_t$, which, as we shall see, is tantamount
to an infinite number of walkers. More explicitly, the problem is that
if one takes the time average of Eq.~(\ref{eq.power.method}) and if
the fluctuations of the $c_{t+1}$ are correlated with
$\ket{\uunder^{(t)}}$ or$\ket{\uunder^{({t+1})}}$ one does not produce
the same state on the left- and right-hand sides of the time-averaged
equation and therefore the time-averaged state need not satisfy the
eigenvalue equation.  The resulting bias has been discussed in the
various contexts.\cite{Hetherington,NBPRB33,UNR}

One way to define a stochastic process is to rewrite the
power method iteration Eq.~(\ref{eq.power.method}) as
\begin{equation}
u^{({t+1})}(\Sp)={1 \over c_{t+1}}\sum_{\S} P(\Sp|\S) g(\S) u^{(t)}(\S),
\label{factor.tm}
\end{equation}
where
\begin{equation}
g(\S)=\sum_{\Sp} G(\Sp|\S)$ and $P(\Sp|\S)=G(\Sp|\S)/g(\S).
\label{DP}
\end{equation}
This is in fact what how transfer matrix \MC\ is defined.  Referring
the reader back to the discussion of Eq.~(\ref{eq.gP}), we note that
in diffusion \MC\ the weight $D$ is defined so that it is not just a
function of the initial state $\S$, but also depends on the final
$\Sp$.  The algorithm given below can trivially be generalized to
accommodate this by making the substitution $g(\S)\to g(\Sp|\S)$.

Equation~(\ref{factor.tm}) describes a process represented by a Monte
Carlo run which, after a few initial equilibration steps, consists of
a time series of $M_0$ updates of all walkers at times labeled by
$t=\ldots,0,1,\ldots,M_0$.  The update at time $t$ consists of two
steps designed to perform stochastically the matrix multiplications in
Eq.~(\ref{factor.tm}).  Following Nightingale and Bl\"ote,
\cite{NBPRB33} the process is defined by the following steps.  Let us
consider one of these updates, the one that transforms the generation
of walkers at time $t$ into the generation at time $t+1$.  We
denote variables pertaining to times $t$ and $t+1$ respectively by
unprimed and primed symbols.
\begin{enumerate}
\item \label{step1}
Update the old walker $(\S_{i},w_{i})$ to yield a temporary
walker $(\Sp_{i},w'_{i})$ according to the transition
probability $P(\Sp_{i}|\S_{i})$, where
$w'_i=g(\S_i)w_i/c'$, for $i =1,\ldots ,N_t$.
Step two, defined below,
can change the number of walkers. To maintain their number
close to a target number, say $N_0$, choose $c' =\hat
\lambda_0(N_t/N_0)^{1/s}$, where $\hat \lambda_0$ is a running estimate
of the eigenvalue $\lambda_0$ to be calculated, where $s\ge 1$ [see
discussion after Eq.~(\ref{power.p})].
\item \label{step2}
{}From the temporary walkers construct the new generation of walkers
as follows
\begin{enumerate}
\item \label{step2.1}
Split each walker $(\Sp,w')$ for which $w'>b_{\/\rm u}$ into two walkers
$(\Sp,{1\over2}w')$.  The choice $b_{\/\rm u}=2$ is a reasonable one.
\item \label{step2.2}
Join pairs $(\Sp_{i},w'_{i})$ and $(\Sp_{j},w'_{j})$
with $w'_{i}<b_{\/\rm l}$ and $w'_{j}<b_{\/\rm l}$ to produce
a single walker $(\Sp_{k},w'_{i}+w'_{j})$, where
$\Sp_{k}=\Sp_{i}$ or $\Sp_{k}=S'_{j}$ with relative
probabilities $w'_{i}$ and $w'_{j}$.  Here $b_{\/\rm l}=1/2$
is reasonable.
\item \label{step2.3}
Walkers for which $b_{\/\rm l}<w'_{i}<b_{\/\rm u}$,
or left single in step~\ref{step2.2}, become members of the new generation
of walkers.
\end{enumerate}
\end{enumerate}
Note that, if the weights $g(\S)$ fluctuate on average more than by a factor of
two, multiple split and join operations may be needed.

It may help to explicate why wildly fluctuating weight adversely
impact the efficiency of the algorithm.  In that case, some walkers
will have multiple descendents in the next generations, whereas others
will have none.  This leads to an inefficient algorithm since any
generation will have several walkers that are either identical or are
closely related, which will produce strongly correlated contributions
to the statistical time averages.  In its final analysis, this is the
same old problem that we encountered in a single-thread algorithm,
where averages would be dominated by few terms with relatively large,
explicitly given statistical weights.
Branching mitigates this problem since walkers that are descendants of
a given walker eventually decorrelate,
but, as discussed in Sections~\ref{sec.singlethread} and \ref{sec.dmc},
the best cure is importance sampling and in practice both strategies are
used simultaneously.

The algorithm described above was constructed so that for any given
realization of $|{\underline u}^{({t})}\rangle$, the expectation value
of $c_{t+1} |{\underline u}^{({t+1})}\rangle$, in accordance with
Eq.~(\ref{eq.power.method}), satisfies
\begin{equation}
{\rm E} \left[ c_{t+1} |{\underline u}^{({t+1})}\rangle\right]=
G |{\underline u}^{({t})}\rangle,
\label{power.1}
\end{equation}
where ${\rm E}(\cdot)$ denotes the conditional average over the
transitions defined by the preceding stochastic process.  More generally,
by $p$-fold iteration one finds \cite{NBPRL60}
\begin{equation}
{\rm E} \left[\left(\prod^p_{b=1} c_{t+b} \right) |{\underline
u}^{({t+p})}\rangle\right]=  G^p |{\underline
u}^{({t})}\rangle. \label{power.p}
\end{equation}

The stationary state average of $|{\underline u}^{({t})}\rangle$ is close to the
dominant eigenvector of $G$, but, as mentioned above, it has
a systematic bias, proportional to $1/N_t$, when the number $N_t$ of walkers is finite.
If, as is the case in some applications, this bias exceeds the statistical
errors, one can again rely on the power method to reduce this bias by
increasing $p$.  If that is done, one is back to the old problem of
having to average products of fluctuating weights, and, as usual, the
variance of the corresponding estimators increases as their bias
decreases.  Fortunately, in practice the population control bias of
the stationary state is quite small, if at all detectable, but even in
those cases, expectation values involving several values of $p$ should
be computed to verify the absence of population control bias.  The
reader is referred
to Refs. \onlinecite{TMreview,Hetherington,UNR,NBPRL60,Runge92b} for a
more detailed discussion of this problem.  Suffice it to mention here,
first of all, that $s$, as defined in the first step of the branching
algorithm given above
is the expected number of time steps it takes to
restore the number of walkers to its target value $N_0$ and, secondly,
that strong population control ($s=1$) tends to introduce a stronger
bias than weaker control ($s>1$).

With Eq.~(\ref{power.p}) one constructs an estimator \cite{NBPRL60}
of the dominant eigenvector $|u^{({\infty})}\rangle$ of the evolution
operator $G$:
\begin{equation}
|\check u^{(p)}\rangle= {1\over M_0}\sum^{M_0}_{t=1}\left(
\prod^{p-1}_{b=0} c_{t-b} \right) |{\underline u}^{({t})}\rangle.
\label{u.estimator}
\end{equation}

For $p=0$, in which case the product over $b$ reduces to unity, this
yields the stationary state of the branching \MC\, which frequently is
treated as the dominant eigenstate of $G$.

Clearly, this branching \MC\
algorithm can be used to compute the right-projected mixed estimators
that were denoted by $X_{\TT}^{(0,\infty)}$ in
Section~\ref{sec.singlethread}.  For this purpose one sets $p'=0$ in
Eq.~(\ref{eq.general.estimator}) and makes the substitution
$\bra{u_\alpha}| = \bra{\uT}|$ and $G^p\ket{u_\beta}| = \ket{{\check
u}^{(p)}}$.  Expressions for several values of $p$ can be computed
simultaneously and virtually for the price of one.
We explicitly mention the important special case
obtained by choosing for the general operator $X$ the evolution
operator $G$ itself.  This yields the following estimator for the
dominant eigenvalue $\lambda_0$ of $G$:
\begin{equation}
\lambda_0 \approx {\sum_{t=1}^{M_0} \left( \prod^{p}_{b=0}c_{t-b}
\right) W^{(t)} \over \sum_{t=1}^{M_0} \left( \prod^{p-1}_{b=0}c_{t-b}
\right) W^{(t-1)}},
\label{growth.estimator}
\end{equation}
where
\begin{equation}
W^{(t)}=\sum_{i=1}^{N_t}w_i^{(t)} \uT(\S_i).
\end{equation}
In diffusion \MC\ this estimator can be used to construct the {\em
growth estimate} of the ground state energy.  That is, since
in that special case $G\approx \exp (-\tau \Ham)$, eigenvalues of
the evolution operator and the Hamiltonian are related by
\beq
E_0=-{1\over \tau} \ln \lambda_0.
\eeq

Besides expressions such as Eq.~(\ref{growth.estimator}) one can construct expresssions with
reduced variance.  These involve the configurational eigenvalue of $G$ or
$\H$ in the same way this was done in our discussion of the single-thread
algorithm. 

Again, in practical applications it is important to combine the raw
branching algorithm defined above with importance sampling.
Mathematically, this works in precisely the same way as in Section
\ref{sec.singlethread} in that one reformulates the same algorithm
in terms of the similarity transform $\Ghat$ with $\ug=\uT$ chosen to
be an accurate, approximate dominant eigenstate [see Eq.~(\ref{eq.Ghatg})].
In the single-thread algorithm, the result is that the fluctuations of
the weights $g$ and their products are reduced.  In the context of the
branching algorithm, this yields reduced fluctuations in the weight
of walkers individually and in the size of the walker population.  One
result is that the population control bias is reduced.  If we ignore
this bias, a more fundamental difference is that the steady state of
the branching algorithm is modified.  That is, in the raw algorithm
the walkers sample the dominant eigenstate of $G$, {\it i.e.,} $\psi_0(\S)$,
but, if the trial state $\ket{\uT}$ is used for importance sampling,
the distribution is $\uT(\S)\psi_0(\S)$, which, of course, is simply
the dominant eigenstate of $\Ghat$.

So far, we have only discussed how mixed expectation values can be
computed with the branching \MC\ algorithm and, as was mentioned
before, this yields the desired result only if one deals with
operators that commute with the evolution operator $G$.  This
algorithm can, however, also be used to perform power method
projections to the left.  In fact, most of the concepts discussed in
Sections \ref{section.commute},\ref{section.diag}, and
\ref{section.non-diag} can be implemented straightforwardly.  To
illustrate this point we shall show how one can compute the left- and
right-projected expectation value of a diagonal operator $X$. Since
the branching algorithm is designed to explicitly perform the
multiplication by $G$ including all weights, all that is required is
the following generalization\cite{Kalospri}, called {\it forward} or
{\it future walking}.

Rather than defining a walker to be the pair formed by a state
and a weight, for forward walking we define the walker to be of the
form $[\S,w,X(\S_{-1}),...,X(\S_{-p'})]$, where
$\S_{-1},\S_{-2},\dots$ are previous states of the walker.  In other
words, each walker is equipped with a finite stack of depth $p'$ of
previous values of the diagonal operator $X$.  In going from one
generation of walkers to the next, the state and weight of a walker
are updated just as before to $\Sp$ and $w'$.  The only new feature is
that the value $X(\S)$ is pushed on the stack:
$[\S,w,X(\S_{-1}),...,X(\S_{-p'})] \to
[\Sp,w',X(\S),X(\S_{-1}),...,X(\S_{-p'+1})]$. In this way, the $p'$
times left-projected expectation value of $X$ is obtained simply by
replacing $X(\S)$ by $X(\S_{-p'})$.  Note that one saves the history
of $X$ rather than a history of configurations only for the purpose of
efficient use of computer memory.

\section{Diffusion Monte Carlo}
\label{sec.dmc}
\subsection{Simple Diffusion Monte Carlo Algorithm}

The diffusion \MC\ method
\cite{GrimmStorer,Anderson7576,CeperleyAlder80,Reynolds82,Moskowitz82}
discussed in this section is an example of the general class of
algorithms discussed in this chapter, all of which rely on stochastic
implementation of the power method to increase the relative spectral
weight of the eigenstates of interest.  In the quantum mechanical context
of the current section, the latter is usually the ground state.  In
this general sense, there is nothing new in this section.  However, a
number of features enter that lead to technical challenges
that cannot be dealt with in a general setting.
There is the problem that the projector,
the evolution operator $G=\exp (-\tau \Ham)$, is only known in the limit
of small imaginary time $\tau$, and in addition, for applications to
electronic structure problems, there are specific
practical difficulties associated with nodes and cusps in the wave
functions.
Bosonic ground states are simpler in that they lack nodes, and we shall
deal with those systems only in passing.  As in the rest of this chapter,
we deal only with the basic concepts.  In particular, we focus on
considerations relevant to the design of an efficient \dMC\ algorithm.
The reader is referred
to Refs.\onlinecite{CeperleyMitas95,Anderson.fn.95,HammondLesterReynoldsbook}
for applications and other issues not covered here.

For quantum mechanical problems, the power method iteration
Eq.~(\ref{eq.power.method}) takes the form
\beq
\ket{\psi(t+\tau)} = e^{-({\cal H}-\ET)\tau} \; \ket{\psi(t)}.
\eeq
Here $\ET$ is a shift in energy such that $E_0-\ET \approx 0$, where
$E_0$ is the ground state energy. In the real-space representation we
have
\beq
\psi(\Rp,t+\tau)
= \intdR \; \bra{\Rp}|e^{-({\cal H}-\ET)\tau}\ket{\R} \; \bra{\R}\ket{\psi(t)}
= \intdR \; \Grprt \; \psi(\R,t)
\label{integGimp},
\eeq
In practical \MC\ settings, the shift $\ET$ is
computed on the fly and consequently is a slowly varying, nearly
constant function of time, but for the moment we take it to be
constant.  The wavefunction $\psi(\R,t)$ is the solution to the
Schr\"odinger equation in imaginary time
\beq
-{1 \over 2 \mu}\nabla^2 \psirt + [{\cal V}(\R)-\ET]
\psirt = - {\partial \psirt \over \partial t}.
\label{schrod.eq}
\eeq
To make contact with Eq.~(\ref{eq.gP}) we should factor the Green
function into a probability conserving part and a weight.  For small
times, this can be accomplished by the following approximation. If, on
the left-hand side, we just had the first term, Eq.~(\ref{schrod.eq})
would reduce to a diffusion equation, whence the method gets its name.
For diffusion, the Green function is probability conserving and is given by
\beq
\Prprt = {e^{-\mu (\Rvecp-\Rvec)^2 / 2\tau} \over (2\pi\tau/\mu)^{3n/2}}
\label{diffusion}.
\eeq
If, on the other hand, we just had the second term then
Eq.~(\ref{schrod.eq}) would reduce to a rate equation for which the
Green function is $\delta(\Rp-\R)$ with prefactor
\beq
\grprt = e^{\tau\{\ET-[\VR+\VRp]/2\}}.
\label{eq.grprt}
\eeq

In combining these ingredients, we have to contend with the following
general relation for noncommuting operators ${\cal H}_i$
\beq
e^{({\cal H}_1+{\cal H}_2)\tau}
= e^{{1\over 2}{\cal H}_1\tau} e^{{\cal H}_2\tau}
e^{{1\over 2}{\cal H}_1\tau} \;+\; {\cal O} (\tau^3)
= e^{{\cal H}_1\tau} e^{{\cal H}_2\tau} \;+\; {\cal O} (\tau^2).
\label{eq.exph1h2}
\eeq

As long as one is dealing with a $\delta$-function, the weight in
Eq.~(\ref{eq.grprt}) is evaluated always at $\Rvecp=\Rvec$, and
therefore the expression on the right can be written also in
non-symmetric form.  However, Eq.~(\ref{eq.exph1h2}) suggests that
the exponent in Eq.~(\ref{eq.grprt}) should be used in symmetric
fashion as written.  This is indeed the form we shall employ for the
time being, but we note that the final version of the diffusion
algorithm employs a nonsymmetric split of the exponent, since
proposed moves are not always accepted.
Since there are other sources of ${\cal O}(\tau^2)$
contributions anyway this does not degrade the performance of the algorithm.
Combination of the preceding ingredients yields
the following approximate, short-time Green function for
Eq.~(\ref{schrod.eq})
\begin{eqnarray}
\Grprt &=& {e^{-\mu (\Rvecp-\Rvec)^2 / 2\tau} \over (2\pi\tau/\mu)^{3n/2}}
\; e^{\tau\{\ET-[\VRp+\VR]/2\}}
\;+\;{\cal O}(\tau^3) \nonumber\\
&=& {e^{-\mu (\Rvecp-\Rvec)^2 / 2\tau} \over (2\pi\tau/\mu)^{3n/2}}
\; e^{\tau\left[\ET-\VR\right]}
\;+\;{\cal O}(\tau^2)
\label{Green.DB}
\end{eqnarray}

For this problem, the power method can be implemented by \MC\
by means of both the single-thread scheme discussed in
Section~\ref{sec.singlethread} and the branching algorithm of
Section~\ref{sec.branching}.  We shall use the latter option in our
discussion.  In this case, one performs the following steps.  Walkers
of the zeroth generation are sampled from $\psiRz = \psiTr$ using the
Metropolis algorithm.  The walkers are propagated forward an imaginary
time $\tau$ by sampling a new position $\Rp$ from a multivariate Gaussian
centered at the old position $\R$ and multiplying the weights of the
walkers by the second factor in Eq.~(\ref{Green.DB}).
Then the split/join step of the branching algorithm is performed to obtain the
next generation of walkers, with weights in the targeted range.

\subsubsection{Diffusion \MC\ with Importance Sampling}
For many problems of interest, the potential energy $\VR$ exhibits
large variations over coordinate space and in fact may diverge at
particle coincidence points.  As we have discussed in the general
case, the fluctuations of weights $g$, produce noisy statistical
estimates. As described in Sections~\ref{sec.singlethread} and
\ref{sec.branching}, this problem can be greatly mitigated by applying
the similarity (or importance sampling) transformation~\cite{GrimmStorer,KLV74}
to the evolution operator.  Employing the general mathematical identity
$S \exp(-\tau \Ham) S^{-1} = \exp (-\tau S \Ham S^{-1})$, this transformation
 can be applied conveniently to the Hamiltonian.  That is, given a
trial function $\psiTr$ one can introduce a distribution $\frt
= \psiTr \psirt$. If $\psirt$ satisfies the Schr\"odinger
equation [Eq.~(\ref{schrod.eq})], it is a simple calculus exercise to
show that $f$ is a solution of the
equation\cite{Reynolds82,Moskowitz82}
\begin{eqnarray}
&&\psiTr ({\cal H}-\ET) \psiTr^{-1} \frt=\nonumber \\
&&-{1 \over 2\mu}\Grad^2 \frt + \Grad\cdot[\Vvecr \frt]
-S(\Rvec) \frt = -{\partial \frt \over \partial t}.
\label{feq}
\end{eqnarray}
Here the {\it velocity} $\Vvec$ is a {\em function} (not an operator),
often referred to
in the literature as the quantum force, and is given by
\beq
\Vvecr = (\vvec_1,\dots,\vvec_n) = {1\over \mu}{\Grad\psiTr\over\psiTr}.
\eeq
The coefficient of the source term, which is responsible for
branching in the diffusion \MC\ context, is
\beq
S(\Rvec)=\ET-\ELr,
\label{source}
\eeq
which is defined in terms of the {\it local energy}
\beq
\ELr = {{\Ham} \psiTr \over \psiTr} = -{1\over 2\mu}{\Grad^2\psiTr\over \psiTr} +
{\cal V}(\Rvec),
\label{ELal}
\eeq
the equivalent of the configurational eigenvalue introduced in
Eq.~(\ref{eq.XT}).

Compared to the original Schr\"odinger equation, to which of course it
reduces for the case $\psiT\equiv 1$, the second term
in Eq.~(\ref{feq}) is new, and corresponds to drift.  Again, one can explicitly
write down the Green function of the equation with just a single term
on the left-hand side.  The drift Green function $\GD$ of the
equation obtained by suppressing all but this term is
\beq
\GDrprt = \delta[\Rvecp-\Rvect(\tau)]
\eeq
where $\Rvect(\tau)$ satisfies the differential equation
\beq
{d\Rvect \over dt} = \Vvecrt
\eeq
subject to the boundary condition $\Rvect(0)=\Rvec$.  Again, at short
times the noncommutativity of the various operators on the left-hand
side of the equation can be ignored and thus one obtains the
following short-time Green function.
\begin{eqnarray}
\Grprt &=&
\intdRpp \;
\delta\left[\Rvecpp-\Rvect(\tau)\right] \;
{e^{-\mu (\Rvecp-\Rvecpp)^2  / 2\tau} \over (2\pi\tau/\mu)^{3n/2}}\;
e^{\tau\{\ET-[\ELrp+\ELr]/2\}}
\;+\;{\cal O}(\tau^2)
\nonumber\\
&=&
{e^{-\mu [\Rvecp-\Rvect(\tau)]^2 / 2\tau} \over (2\pi\tau/\mu)^{3n/2}}\;
e^{\tau\{\ET-[\ELrp+\ELr]/2\}}
\;+\;{\cal O}(\tau^2).
\label{Green.DDB}
\end{eqnarray}

Eq.~(\ref{Green.DDB}) again can be viewed in our general
framework by defining the probability conserving generalization of
Eq.~(\ref{diffusion})
\beq
\Prprt = {e^{-\mu [\Rvecp-\Rvect(\tau)]^2 / 2\tau}
\over (2\pi\tau/\mu)^{3n/2}}
\label{driftdiffusion}
\eeq
and the remainder of the Green function is
$\delta[\Rvecp-\Rvect(\tau)]$ with prefactor
\beq
\grprt= e^{\tau\{\ET-[\ELrp+\ELr]/2\}},
\label{eq.grprt'}
\eeq
which is the analog of Eq.~(\ref{eq.grprt}).

When employed for the branching \MC\ algorithm, the factorization
given in Eqs.~(\ref{driftdiffusion}) and (\ref{eq.grprt'}) differs from the
original factorization Eqs.~(\ref{diffusion}) and (\ref{eq.grprt}) in
two respects: (a) the walkers do not only diffuse but also drift towards
the important regions, {\it i.e.,} in the direction in which $\vert\psiTr\vert$
is increasing; and (b) the branching term is better behaved since it
depends on the local energy rather than the potential.
In particular, if the trial function $\psiTr$ obeys cusp conditions~\cite{Kato}
then the local energy at particle coincidences is finite even though the
potential may be infinite.  If one were so
fortunate that $\psiTr$ is the exact ground state, the branching factor
would reduce to a constant, which can be chosen to be unity by choosing
$\ET$ to be the exact energy of that state.

The expressions, as written, explicitly contain the expression
$\Rvect(\tau)$, which has to be obtained by integration of the velocity
$\Vvecr$. Since we are using a short-time expansion anyway, this exact
expression may be replaced by the approximation
\beq
\Rvect(\tau) = \Rvec + \Vvecr \tau +{\cal O}(\tau^2).
\label{eq.Rtau}
\eeq
In the improvements discussed below, designed to reduce
time-step error, this expression is
improved upon so that regions where $\Vvec$ diverges do not make large
contributions to the overall time-step error.

\subsubsection{Fixed-Node Approximation}
\label{sec.fixednode}

The absolute ground state of a Hamiltonian with particle exchange
symmetry is bosonic.  Consequently, unless one starts with a
fermionic, {\it i.e.,} antisymmetric wavefunction and implements the
power method in a way that respects this antisymmetry, the power
method will reduce the spectral weight of the fermionic eigenstate to
zero relative to the weight of the bosonic state.  The branching
algorithm described above assumes all weights are positive and
therefore is incompatible with the requirement of preserving
antisymmetry.  The algorithm needs modification, if we are interested
in the fermionic ground state.

If the nodes of the fermionic ground state were known, they could be
imposed as boundary conditions~\cite{Anderson7576} and the problem
could be dealt with by solving the Schr\"odinger equation within a
single, connected region bounded by the nodal surface, a region we
shall refer to as a {\it nodal pocket.}  Since all the nodal pockets
of the ground state of a fermionic system are equivalent, this would
yield the exact solution of the problem everywhere.  Unfortunately,
the nodes of the wavefunction of an $n$-particle system form a
$(3n-1)$-dimensional surface, which should not be confused with the
nodes of single-particle orbitals.  Of this full surface, in general,
only the $(3n-3)$-dimensional subset, corresponding to the coincidence
of two like-spin particles, is known.  Hence, we are forced to employ
an approximate nodal surface as a boundary condition to be satisfied
by the solution of the Schr\"odinger equation.  This is called the
{\it fixed-node} approximation.  Usually, one chooses for this purpose
the nodal surface of an optimized trial wave
function~\cite{Anderson7576}, and such nodes can at times yield a very
accurate results if sufficient effort is invested in optimizing the
trial wavefunction.

Since the imposition of the boundary condition constrains the
solution, it is clear that the fixed-node energy is an upper bound on
the true fermionic ground state energy.  In diffusion \MC\
applications, the fixed-node energy typically has an error which is
five to ten times smaller than the error of the variational energy
corresponding to the same trial wavefunction, though this can
vary greatly depending on the system and the trial wavefunction
employed.

For the \MC\ implementation of this approach one has to use an
approximate Green function, which, as we have seen, may be obtained by
iteration of a short time approximation.  To guide the choice of an
approximant accurate over a wide time range, it is useful to consider
some general properties of the fixed-node approximation.
Mathematically, the latter amounts to solution of the Schr\"odinger
equation in a potential that equals the original physical potential
inside the nodal pocket of choice, and is infinite outside.  The
corresponding eigenstates of the Hamiltonian are continuous and
vanish outside the nodal pocket.  Note that the solution can be
analytically continued outside the initial nodal pocket only if the
nodal surface is exact, otherwise there is a derivative discontinuity at
the nodal surface.  These properties are shared by the Green
function consistent with the boundary conditions.  This can be seen by
writing down the spectral decomposition for the evolution operator in
the position representation
\beq
\Grprt = \bra{\Rvecp}|e^{-\tau \Ham}\ket{\Rvec} =
\sum_i \psi_i(\Rvecp) e^{-\tau E_i} \psi_i(\Rvec)
\label{eq.Gspectral}
\eeq
where the $\psi_i$ are the eigenstates satisfying the required boundary
conditions.  For notational convenience only, we have assumed that the
spectrum has a discrete part only and that the wavefunctions can be
chosen to be real.  The Green function vanishes
outside the nodal pocket and generically vanishes linearly at the nodal
surface, as do the wavefunctions.

The Green function of interest in practical applications is
the one corresponding to importance sampling, the similarity transform
of Eq.~(\ref{eq.Gspectral})
\beq
\Ghatrprt =
\psiTrp \; \bra{\Rvecp}|e^{-\tau \Ham}\ket{\Rvec}{1\over \psiTr}
\label{eq.Gspectrali}
\eeq
This Green function vanishes at the nodes quadratically in its first
index, which, in the \MC\ context is the one that determines to which
state a transition is made from a given initial state.

The approximate Green functions of Eqs.~(\ref{Green.DB}) and
(\ref{Green.DDB}) have tails that extend beyond the nodal surface and,
consequently, walkers sampled from these Green functions have a finite
probability of attempting to cross the node.  Since expectation values
ought to be calculated in the $\tau \to 0$ limit the relevant quantity
to consider is what fraction of walkers attempt to cross the node {\it
per unit time} in the $\tau \to 0$ limit.  If the Green function of
Eq.~(\ref{Green.DB}) is employed, this is a finite number, whereas, if
the importance-sampled Green function of Eq.~(\ref{Green.DDB}) is
employed, no walkers cross the surface since the velocity $\Vvec$ is
directed away from the nodes and diverges at the nodes.  In practice,
of course, the calculations are performed for finite $\tau$, but the
preceding observation leads to the conclusion that in the former case it
is necessary to reduce to zero the weight of a walker that has crossed
a node, {\it i.e.,} to kill the walker, while in the latter case one
can either kill the walkers or reject the proposed move, since in the
$\tau \to 0$ limit they yield the same result.

We now argue that, for finite $\tau$, rejecting moves is
the choice with the smaller time-step error
when the importance-sampled Green function is employed.
Sufficiently close to a node, the component of the velocity perpendicular
to the node dominates all other terms in the Green function and
it is illuminating to consider a free particle in one dimension
subject to the boundary condition that $\psi$ have a node at $x=0$.
The exact Green function for this problem is
\beq
G(x',x,\tau) = {1 \over \sqrt{2\pi\tau/\mu}}
[e^{-\mu(x'-x)^2/2\tau} - e^{-\mu(x'+x)^2/2\tau}],
\label{G.1d}
\eeq
while the corresponding importance sampled Green function is
\beq
{\hat G}(x',x,\tau) = {x' \over x \sqrt{2\pi\tau/\mu}}
[e^{-\mu(x'-x)^2/2\tau} - e^{-\mu(x'+x)^2/2\tau}].
\label{G.1di}
\eeq
We note that the integral of the former, over the region $x>0$, is less
than one and decreases with time.  In terms of the usual language used
for diffusion problems, this is because of absorption at the $x=0$
boundary.  In our case, this provides the mathematical justification
for killing the walkers that cross.  On the other hand, the integral
of the Green function of Eq.~(\ref{G.1di}) equals one.  Consequently,
for finite $\tau$ it seems likely that algorithms that reject moves
across the node, such as the one discussed in
Section~\ref{sec.Improved} yield a better approximation than
algorithms that kill the walkers that cross the node.

As mentioned, it can be shown that all the nodal pockets of the ground
state of a fermionic system are equivalent and trial wavefunctions
are constructed to have the same property.  Consequently, the \MC\
averages will not depend on the initial distribution of the walkers
over the nodal pockets.  The situation is more complicated for excited
states, since different nodal pockets of excited-state wavefunctions
are not necessarily equivalent, neither for bosons nor for fermions.
Any initial state with walkers distributed randomly over nodal pockets
pockets, will evolve to a steady state distribution with walkers only
in the pocket with the lowest average local energy, at least if we
ignore multiple-node crossings and assume a sufficiently large number
of walkers, so that fluctuations in the average local energy can be
ignored.

\subsubsection{Problems with Simple Diffusion Monte Carlo}
\label{sec.simpledmcprobs}
The \dMC\ algorithm corresponding to Eq.~(\ref{Green.DDB}) is in fact
not viable for a wavefunction with nodes for the following two
reasons.  Firstly, in the vicinity of the nodes the local energy of
the trial function $\psiT$ diverges inversely proportional to the
distance to the nodal surface.  For nonzero $\tau$, there is a nonzero
density of walkers at the nodes.  Since the nodal surface for a system
with $n$ electrons is $3n-1$ dimensional, the variance of the local
energy diverges for any finite $\tau$. In fact, the expectation value
of the local energy also diverges, but only logarithmically.
Secondly, the velocity of the electrons at the nodes diverges
inversely as the distance to the nodal surface.  The walkers that are
close to a node at one time step, drift at the next time step to a
distance inversely proportional to the distance from the node.  This
results in a charge distribution with a component that falls off as
the inverse square of distance from the atom or molecule, whereas in
reality the decay is exponential.  These two problems are
often remedied by introducing
cut-offs in the values of the local energy and the
velocity\cite{DePasquale88,GarmerAnderson88}, chosen such that they
have no effect in the $\tau \to 0$ limit, so that the results
extrapolated to $\tau = 0$ are correct.  In the next section better
remedies are presented.

\subsection{Improved Diffusion Monte Carlo Algorithm}
\label{sec.Improved}
\subsubsection{The limit of Perfect Importance Sampling}
\label{sec.acc-rej}
In the limit of perfect importance sampling, that is if
$\psiTr=\psizr$, the energy shift $\ET$ can be chosen such that the
branching term in Eq.~(\ref{feq}) vanishes identically for all ${\bf
R}$.  In this case, even though the energy can be obtained with zero
variance, the steady state distribution of the \dMC\ algorithm discussed
above is only approximately the desired distribution $\psiT^2$,
because of the time-step error in the Green function.  However, since
one has an explicit expression, $\psiT^2$, for the distribution to be
sampled, it is possible to use the Metropolis algorithm, described in
Section~\ref{sec.Metropolis}, to sample the desired distribution
exactly.  Although the ideal $\psiTr=\psizr$ is never achieved in
practice, this observation leads one to devise an improved algorithm
that can be used when moderately good trial wavefunctions are known.

If for the moment we ignore the branching term in
Eq.~(\ref{feq}), then we have the equation
\begin{equation}
-{1\over 2\mu}\Grad^2 f + \Grad\cdot(\Vvec f) =
-{\partial f \over \partial t}
\label{vmc}.
\end{equation}
This equation has a known steady-state solution $f = \psiT^2$ for
any $\psiT$, which in the limit of perfect importance sampling is the
desired distribution.
However, the approximate drift-diffusion Green
function used in the \MC\ algorithm defined by Eq.~(\ref{Green.DDB})
without the branching factor,
is not the exact Green function of Eq.~(\ref{vmc}).
Therefore, for any finite time step $\tau$,
we do not obtain $\psiT^2$  as a steady state, even in the ideal case.
Following Reynolds \etal ~\cite{Reynolds82}, we can change the algorithm
in such a way that it {\it does} sample
$\psiT^2$ in the ideal case, which also reduces the time-step error
in nonideal, practical situations. This is accomplished by using a
generalized \cite{Hastings,CeperleyChesterKalos,KalosWhitlock}
Metropolis algorithm \cite{Metropolis}.
The approximate drift-diffusion Green function is used to propose moves,
which are then accepted with probability
\begin{equation}
p = {\rm min} \left({\Gtrrpt \psiTrp^2 \over
\Gtrprt \psiTr^2},1\right)\equiv 1-q \label{accep},
\end{equation}
in accordance with the detailed balance condition.

As was shown above, the true fixed-node Green function
vanishes outside the nodal pocket of the trial
wavefunction.  However, since we are using an approximate Green function,
moves across the nodes will be proposed for any
finite $\tau$.  To satisfy the boundary conditions of the fixed-node
approximation these proposed moves are always rejected.

If we stopped here, we would have an exact and efficient variational
\MC\ algorithm to sample from $\psiT^2$.  Now, we
reintroduce the branching term to convert the steady-state
distribution from $\psiT^2$ to $\psiT \psi_0$.  This is
accomplished by reweighting the walkers with the branching factor
[see Eq.~(\ref{Green.DDB})]
\begin{eqnarray}
\Delta w = \left\{
\begin{array}{ll}
\exp\{ {1\over 2}[S(\Rvecp)+S(\Rvec)]\taueff\}\;\;
&\mbox{for an accepted move},
\\
\exp[S(\Rvec)\taueff]
\;\;
&\mbox{for a rejected move}, \end{array} \right.
\label{eq.branch}
\end{eqnarray}
where $S$ is defined in Eq.~(\ref{source}).  An effective time step
$\taueff$, which will be defined presently, is required because the
Metropolis algorithm introduces a finite probability of not moving
forward and rejecting the proposed configuration.
Before defining $\taueff$, we note that
an alternative to
expression~(\ref{eq.branch}) is obtained by replacing the two
reweighting factors by a single expression,
\beq
\Delta w = \exp\left[\left\{{p\over 2}\left(S(\Rvecp)+
S(\Rvec)\right)+qS(\Rvec)\right\}\taueff\right]
\;\;\mbox{for all moves}
\label{eq.branch.pq}
\eeq
This expression, written down with Eq.~(\ref{eq.X.MCav2}) in mind,
yields somewhat smaller fluctuations and time-step error than
expression~(\ref{eq.branch}).

Following Reynolds \etal \cite{Reynolds82},
an effective time step $\taueff$ is introduced in Eq.(~\ref{eq.branch.pq})
to account for the changed rate of diffusion.  We set
\begin{equation}
\taueff = \tau {\langle p\;\Delta R^2 \rangle
\over \langle \Delta R^2\rangle},
\label{taueff}
\end{equation}
where the angular brackets denote the average over all attempted
moves, and $\Delta R$ are the displacements resulting from diffusion.
This equals ${\langle \Delta R^2 \rangle_{\rm accepted} / \langle
\Delta R^2\rangle}$ but again has somewhat smaller fluctuations.

An estimate of $\taueff$ is readily obtained iteratively from sets of
equilibration runs. During the initial run, $\taueff$ is set equal to
$\tau$.  For the next runs, the value of $\taueff$ is obtained from the
values of $\taueff$ computed with Eq.~(\ref{taueff}) during the
previous equilibration run. In practice, this procedure converges in
two iterations, which typically consume less than 2\% of the total
computation time.
Since the statistical errors in $\taueff$ affect the results obtained,
the number of Monte Carlo steps performed during the equilibration phase
needs to be sufficiently large that this is not a major component of the
overall statistical error.

The value of $\taueff$ is a measure of the rate at which the
\MC\ process generates uncorrelated configurations, and thus a
measure of the efficiency of the computation.  Since the acceptance
probability decreases when $\tau$ increases, $\taueff$ has a maximum as
a function of $\tau$. However, since the time-step error increases with
$\tau$, it is advisable to use values of $\tau$ that are
smaller than this ``optimum".

Algorithms that do not exactly simulate the equilibrium distribution
of the drift-diffusion equation if the branching term is suppressed,
\ie, algorithms that do not use the Metropolis accept/reject mechanism,
can for sufficiently large $\tau$ have time-step errors that make the
energy estimates higher than the variational energy.  On the other
hand, if the drift-diffusion terms are treated exactly by including an
accept/reject step, the energy, evaluated for any $\tau$,
must lie below the variational energy,
since the branching term enhances the weights of the low-energy
walkers relative to that of the high-energy walkers.

\subsubsection{Persistent Configurations} \label{sec.persist}
As mentioned above, the accept/reject step has the desirable feature of
yielding the exact electron distribution in the limit that the trial
function is the exact ground state.  However, in practice the trial
function is less than perfect and as a consequence the accept/reject
procedure can lead to the occurrence of persistent configurations, as
we now discuss.

\begin{figure}
\centerline{\psfig{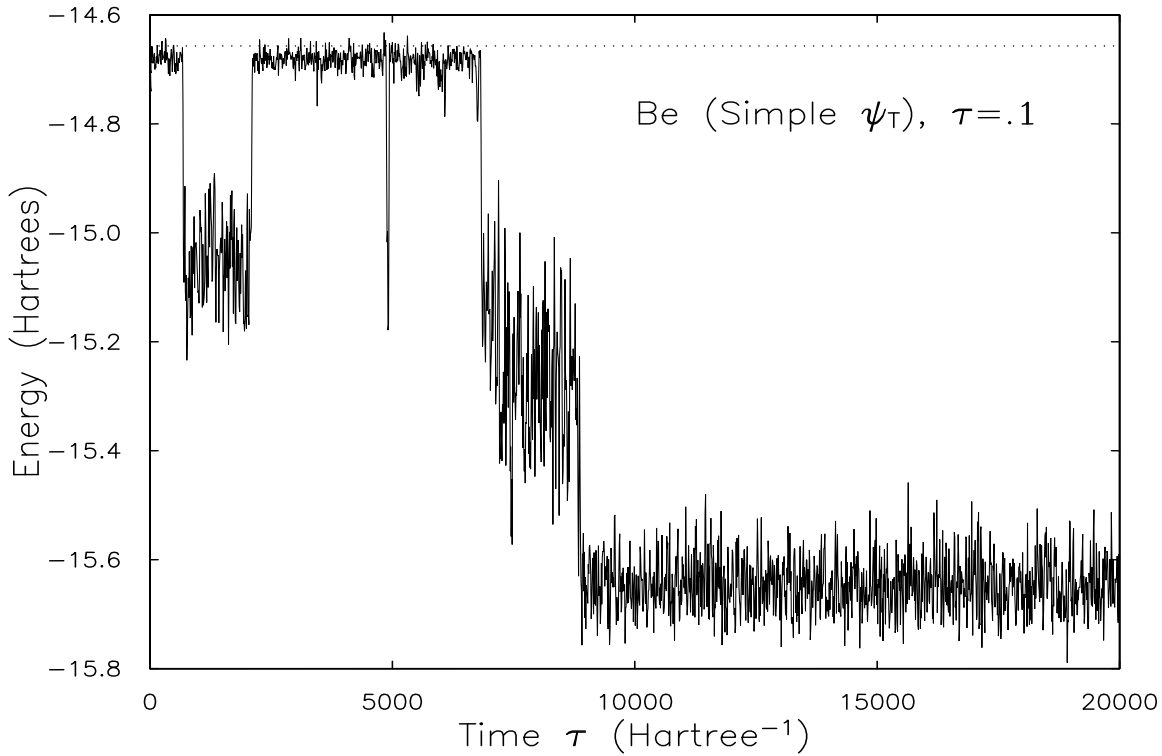}}
\vskip .5cm
\caption[c2]{Illustration of the persistent configuration catastrophe.
The dotted horizontal line is the true fixed-node energy for a
simple Be wavefunction extrapolated to $\tau=0$.}
\label{pl.persist}
\end{figure}

For a given configuration $\Rvec$, consider the quantity $Q=\langle
q\Delta w \rangle$, where $q$ and $\Delta w$ are the rejection
probability and the branching factor given by Eqs.~(\ref{accep}) and
(\ref{eq.branch.pq}).  The average in the definition of $Q$ is over
all possible moves for the configuration $\Rvec$ under consideration.
If the local energy at $\Rvec$ is relatively low and if $\taueff$ is
sufficiently large, $Q$ may be in excess of one.  In that case, the
weight of the walker at $\Rvec$, or more precisely, the total weight
of all walkers in that configuration will increase with time, except
for fluctuations, until the time-dependent trial energy $\ET$ adjusts
downward to stabilize the total population.  This population contains
on average a certain number of copies of the persistent configuration.
Since persistent configurations must necessarily have an energy that
is lower than the true fixed-node energy, this results in a negatively
biased energy estimate.  The persistent configuration may disappear
because of fluctuations, but the more likely occurrence is that it is
replaced by another configuration that is even more strongly
persistent, \ie, one that has an even larger value of $Q=\langle
q\Delta w \rangle$.  This process produces a cascade of configurations
of ever decreasing energies.  Both sorts of occurrences are
demonstrated in Fig.~\ref{pl.persist}.  Persistent configurations are
most likely to occur near nodes, or near nuclei if $\psiT$ does not obey
the cusp conditions.  Improvements to the
approximate Green function in these regions, as discussed in the next
section, help to reduce greatly the probability of encountering
persistent configurations to the point that they are never encountered
even in long \MC\ runs.

Despite the fact that the modifications described in the next section
eliminate persistent configurations for the systems we have studied,
it is clearly desirable to have an algorithm that cannot display this
pathology even in principle.  One possible method is to replace
$\taueff$ in Eq.~(\ref{eq.branch}) by $\tau$ for an accepted move and
by zero for a rejected move.  This ensures that $\Delta w$ never
exceeds unity for rejected moves, hence eliminating the possibility of
persistent configurations.  Further, this has the advantage that it is
not necessary to determine $\taueff$.  Other possible ways to
eliminate the possibility of encountering persistent configurations
are discussed in Ref.~\onlinecite{our.dmc}.

\subsubsection{Singularities} \label{sec.sing}
The number of iterations of Eq.~(\ref{integGimp}) required for the
power method to converge to the ground state grows inversely with the
time step $\tau$.  Thus, the statement made above, \viz\ that the Green
function of Eq.~(\ref{Green.DB}) is in error only to ${\cal O}(\tau^2)$,
would seem to imply that the errors in the electron distribution and
the averages calculated from the short-time Green function are of
${\cal O}(\tau)$.  However, the presence of
non-analyticities and divergences in the local energy
and the velocity may invalidate this argument: the short-time Green
function may lack uniform convergence in $\tau$ over $3n$-dimensional
configuration space.  Consequently, an approximation that is designed
to be better behaved in the vicinity of singularities and therefore
behaves more uniformly over space may outperform an approximation that
is correct to a higher order in $\tau$ for generic points in
configuration space, but ignores these singularities.

Next we discuss some important singularities that one may encounter
and their implications for the \dMC\ algorithm.  The behavior of the
local energy $\EL$ and velocity ${\bf V}$ near nodes of $\psiT$ are
described in Table~\ref{table.singularities}.  Although the true wave
function $\psiz$ has a constant local energy $E_0$ at all points in
configuration space, the local energy of $\psiT$ diverges at most
points of the nodal surface of $\psiT$ for almost all the $\psiT$ that
are used in practice, and it diverges at particle
coincidences (either electron-nucleus or electron-electron) for wave
functions that fail to obey cusp conditions~\cite{Kato}.
The velocity ${\bf
V}$ diverges at the nodes and for the Coulomb potential has a
discontinuity at particle coincidences both for approximate and for
the true wavefunction.  For the nodeless wavefunction of
Lennard-Jones particles in their ground state,\cite{MMN96} ${\bf V}$
diverges as $r^{-6}$ in the inter-particle distance $r$.  Since the
only other problem for these bosonic systems is the divergence of $\EL$ at
particle coincidences, we continue our discussion for
electronic systems and refer the interested reader to the literature\cite{MMN96}
for details.

\begin{table}[htb]
\caption[tab]{\small \parindent .5cm \narrower Behavior of the local
energy $\EL$ and velocity $v$ as a function of the distance $\Rperp$ of an
electron to the nearest singularity. The behavior of various quantities
is shown for an electron approaching a node or another particle, either
a nucleus or an electron.  The singularity in the local energy at
particle overlap is only present for a $\psiT$ that fails to satisfy
the cusp conditions.\par\vspace{5mm}}
\begin{tabular}{c|c|c}
Region & Local energy & Velocity\\
\tableline
Nodes& $\EL \sim \pm {1\over R_\perp}$ for $\psiT$& $v \sim {1\over R_\perp}$\\
     &  $\EL=E_0$ for $\psi_0$&\\
 & \\
Electron- & $\EL \sim {1 \over x}$ for some $\psiT$& $v$ has a discontinuity\\
nucleus/electron   & $\EL=E_0$ for $\psi_0$ & for both $\psiT$ and $\psi_0$\\
\end{tabular}
\label{table.singularities}
\end{table}

\begin{figure}
\centerline{\psfig{figure=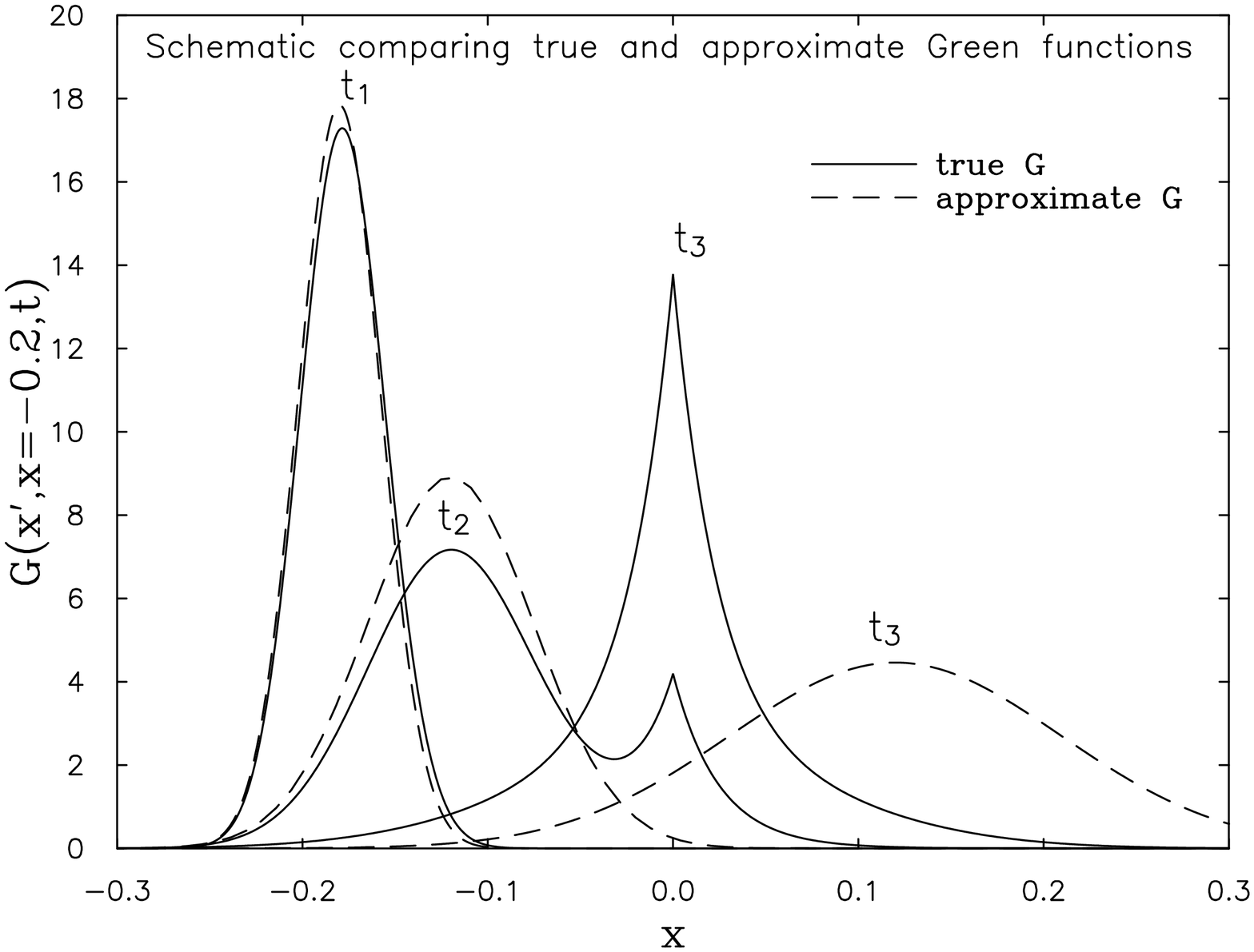,height=3.9truein,width=6.in}}
\vskip .5cm
\caption[c3]{\footnotesize
Schematic comparing the qualitative behaviors of the true $G$
with the approximate $G$ of Eq.~(\ref{Green.DDB}) of an electron that
is located at $x=-0.2$ at time $t=0$ and evolves in the presence of
a nearby nucleus located at $x=0$.  The Green functions are plotted
for three times: $t_1<t_2<t_3$.}
\label{fig.overshoot}
\end{figure}

The divergences in the local energies cause large fluctuations in the
population size: negative divergences lead to large local population
densities and positive divergences lead to small ones.
The divergence of ${\bf V}$ at the nodes typically leads to a proposed next
state of a walker in a very unlikely region of configuration space
and is therefore likely to be rejected.
The three-dimensional velocity of an electron which is close to a
nucleus is directed towards the nucleus.  Hence the true Green function,
for sufficiently long time, exhibits a peak at the nucleus, but
the approximate Green function of Eq.~(\ref{Green.DDB}) cause electrons
to overshoot the nucleus.
This is illustrated in Fig.~\ref{fig.overshoot}, where we show the
qualitative behavior of the true Green function and of the approximate
Green function for an electron which starts at $x=-0.2$ in the presence of a
nearby nucleus at $x=0$.  At short time $t_1$, the approximate
Green function
of Eq.~(\ref{Green.DDB}) agrees very well with the
true Green function.  At a longer time $t_2$, the true Green function
begins to develop a peak at the nucleus which is absent in the approximate
Green function, wheras at a yet longer time $t_3$, the true Green function
is peaked at the nucleus while the approximate Green function
has overshot the nucleus.

The combined effect of these nonanalyticities is a large time-step
error, which can be of either sign in the energy, and large
statistical uncertainty in the computed expectation values.  We now
give a brief description of how these nonanalyticities are
treated.  The
divergence of the local energy at particle coincidences is cured
simply by employing wavefunctions that obey cusp
conditions~\cite{Kato}.  The other nonanalyticities are addressed by
employing a modification of the Green function of Eq.~(\ref{Green.DDB})
such that it incorporates the divergence of $\EL$ and ${\bf V}$
at nodes and the discontinuity in ${\bf V}$ at particle coincidences but
smoothly reduces to
Eq.~(\ref{Green.DDB}) in the short-time limit or in the limit that the
nearest nonanalyticity is far away.  The details can be found in
Ref.~\onlinecite{our.dmc}.  The modified algorithm has a time-step
error which is two to three orders of magnitude smaller~\cite{our.dmc}
than the simple algorithm corresponding to Eq.~(\ref{Green.DDB})
with cutoffs imposed on $\EL$ and $\Vvec$.

We have used the application to all-electron electronic structure
calculations to illustrate the sort of problems that can lead to large
time-step errors and their solution.  Other systems may exhibit only a
subset of these problems or a modified version of them.  For example,
in calculations of bosonic clusters~\cite{MMN96} there are no nodes to
contend with, while, in electronic structure calculations employing
pseudopotentials~%
\cite{Christiansen,HammondReynoldsLester87,%
FahyWangLouie,FladSavinPreuss92},
or pseudo-Hamiltonians~\cite{PseudoHamiltonian} the potential need not
diverge at electron-nucleus coincidences.  We note, in passing, that
the use of the latter methods has greatly extended the practical
applicability of \qMC\ methods to relatively large systems of
practical interest~\cite{FahyWangLouie,Pseudopotential.Illinois} but
at the price of introducing an additional
approximation~\cite{Pseudopotential.portability}.

\section{Closing Comments}

The material presented above was selected to describe from a unified
point of view \MC\ algorithms as employed in seemingly unrelated areas
in quantum and statistical mechanics.  Details of applications were
given only to explain general ideas or important technical problems,
such as encountered in \dMC.  We ignored a whole body of literature,
but we wish to just mention a few topics.  Domain Green function
\MC\cite{CeperleyKalosbook,SchmidtMoskowitz,KLV74,Ceperley83}
is one that comes very close to topics that were covered.
In this method the Green function is sampled exactly by iterating upon
an approximate Green function.  Infinite iteration, which the \MC\
method does in principle, produces the exact Green function.
Consequently, this method lacks a time-step error, and in this
sense has the advantage of being exact.  In practice, there are other
reasons, besides the time-step error that force the algorithm to move
slowly through state space, and currently the available algorithms
seem to be less efficient than \dMC, even when one accounts for the
effort required to perform the extrapolation to vanishing time step.

Another area that we just touched upon in passing is path integral \MC.\cite{PIMC}
Here we remind the reader that path integral \MC\ is a particularly
appealing alternative for the evaluation of matrix elements such as
such as $\X \alpha \beta {p'} p$ in
Eq.~(\ref{eq.general.estimator}).  The advantage of this method is that
no products of weights appear, but the disadvantage is that it seems to
be more difficult to move rapidly through state space.  This is a
consequence of the fact that branching algorithms propagate just a
single time slice through state space, whereas path integral methods
deal with a whole stack of slices, which for sampling purposes tends to
produce a more rigid object.
Finally, it should be mentioned that we ignored the vast
literature on quantum lattice systems~\cite{QuantumLattice}.

In splitting the evolution operator into weights and probabilities
[see Eq.~(\ref{eq.ghatPhat})] we assumed that the weights were
non-negative.  To satisfy this requirement, the fixed-node
approximation was employed for fermionic systems.  An approximation in
the same vein is the fixed-phase approximation,\cite{FixedPhase} which
allows one to deal with systems in which the wavefunction is
necessarily complex valued. The basic idea here is analogous to
that underlying the fixed-node approximation.  In the latter, a trial
function is used to approximate the nodes while \dMC\ recovers the
magnitude of the wavefunction.  In the fixed phase approximation, the
trial function is responsible for the phase and \MC\ produces the
magnitude of the wavefunction.

\acknowledgments
This work was supported by the (US) National Science Foundation through
Grants DMR-9725080 and CHE-9625498 and by the Office of Naval
Research.  The authors thank Richard Scalettar and Andrei Astrakharchik
for their helpful comments.

\end{document}